\documentclass[prd, twocolumn, superscriptaddress,floatfix, nofootinbib, preprintnumbers]{revtex4-2}
\usepackage[utf8]{inputenc}
\usepackage[colorlinks=true,citecolor=blue]{hyperref}
\usepackage{amssymb}
\usepackage{amsmath}
\usepackage{graphicx}
\usepackage{footmisc}
\usepackage{url}
\usepackage{multirow}
\usepackage{makecell}
\usepackage{hhline}
\usepackage{comment}
\usepackage{array}
\usepackage{float}
\usepackage{ragged2e}
\usepackage{threeparttable}
\usepackage{enumitem}

\usepackage{xcolor}

\def\beq{\begin{equation}}
\def\eeq{\end{equation}}
\newcommand{\bea}{\begin{eqnarray}\begin{aligned}}
\newcommand{\eea}{\end{aligned}\end{eqnarray}}

\newcommand{\mycomment}[1]{}

\begin{document}

\title{Feature Selection with Distance Correlation}

\author{Ranit Das}
\email{ranit@physics.rutgers.edu}
\affiliation{NHETC, Dept.\ of Physics and Astronomy, Rutgers University, Piscataway, NJ 08854, USA}

\author{Gregor Kasieczka}
\email{gregor.kasieczka@uni-hamburg.de}
\affiliation{Institut f\"{u}r Experimentalphysik, Universit\"{a}t Hamburg, 22761 Hamburg, Germany}
\affiliation{Center for Data and Computing in Natural Sciences (CDCS), 22607 Hamburg, Germany}

\author{David Shih}
\email{shih@physics.rutgers.edu}
\affiliation{NHETC, Dept.\ of Physics and Astronomy, Rutgers University, Piscataway, NJ 08854, USA}

\begin{abstract}

Choosing which properties of the data to use as input to multivariate decision algorithms --- a.k.a. feature selection --- is an important step in solving any problem with machine learning. 
While there is a clear trend towards training sophisticated deep networks on large numbers of relatively unprocessed inputs (so-called automated feature engineering), for many tasks in physics, sets of theoretically well-motivated and well-understood features already exist. Working with such features can bring many benefits, including greater interpretability, reduced training and run time, and enhanced stability and robustness.
We develop a new feature selection method based on Distance Correlation (DisCo), 
and demonstrate its effectiveness on the tasks of boosted top- and $W$-tagging. Using our method to select features from a set of over 7,000 energy flow polynomials, 
we show that we can match the performance of much deeper architectures,
by using only ten features and two orders-of-magnitude fewer model parameters.

\end{abstract}

\maketitle

\section{Introduction}

Recently there has been enormous progress in training supervised deep learning classifiers to perform object and event identification at the LHC. Deep learning classifiers that make use of low-level information (such as the four vectors of all the reconstructed particles in a jet or event) have been shown to achieve impressive performance gains over cut-based methods and shallow classifiers trained on high level kinematic features, translating directly into better physics performance~\cite{landscape,1808887,Karagiorgi:2021ngt}.

One very fruitful benchmark task for developing new architectures has been boosted top tagging, i.e.\ classifying jets from hadronic-decays of boosted top quarks against the background of light quark and gluon jets. Boosted top jets have a rich, varied and subtle substructure that deep learning classifiers can leverage and exploit to enhance their performance. Boosted top tagging has been a fertile canvas for working with a wide variety deep learning methods, such as DNNs \cite{ann_constituent,topodnn,nsub_landscape}, CNNs~\cite{Kasieczka:2017nvn,dshih_CNN, irc_cnn},  recurrent \cite{rnn} and recursive NNs \cite{recnn,sebast_recNN}, sets~\cite{Komiske:2018cqr}, graph NNs \cite{particlenet,Dreyer:2020brq,graph_lund,jedinet}, and transformers~\cite{Mikuni:2021pou,part}. Performance gains have also been reported using approaches that exploit the underlying Lorentz invariance \cite{Butter:2017cot,Erdmann:2018shi,Bogatskiy:2020tje,lorenznet,pelican}.

However, all of these high-performing deep learning methods are black boxes, and there has been a parallel effort in AI interpretability / explainability to understand ``what the machine learns" \cite{interpret_1,interpret_2,interpret_3,interpret_4,Agarwal:2020fpt}.
Recently, an important step in this direction came from \cite{ado}, which developed a new forward feature selection technique to efficiently scan through more than 7,000 energy flow polynomials
(EFPs)~\cite{efp} --- i.e. quantities that measure the energy distribution inside a jet ---
in order to identify a small number (typically of order ten) that together reproduce as closely as possible the performance of a state-of-the-art black-box NN classifier. Their method relied on a score called ``average decision ordering" (ADO) which measures how often a given feature has the same decision ordering as the reference classifier. This method has been applied to W-jets \cite{ado}, muons \cite{ado_muons}, electrons \cite{ado_electrons}, and semi-visible dark-jets \cite{ado_smj}. 

Aside from shedding light on ``what the machine learns", constructive feature selection methods 
can have several other interesting applications. 
Classifiers based on high-level features (HLFs) could be more robust against domain shifts and more easy to calibrate with collider data (as a smaller number of distributions need to be validated).
Also, a classifier trained on only a few inputs could be made much more lightweight (far fewer parameters), leading to less intensive training and faster evaluation time. This could have important applications to ML with microsecond inference times, e.g.\ for the LHC trigger. Finally, even if attempting to replicate a state-of-the-art deep learning classifier with a set of HLFs falls short, it might have important physics implications, as it could teach us that the set of HLFs being used is incomplete and does not fully capture all the correlations in the data. 

\begin{figure*}
    \centering
    \includegraphics[width=\linewidth]{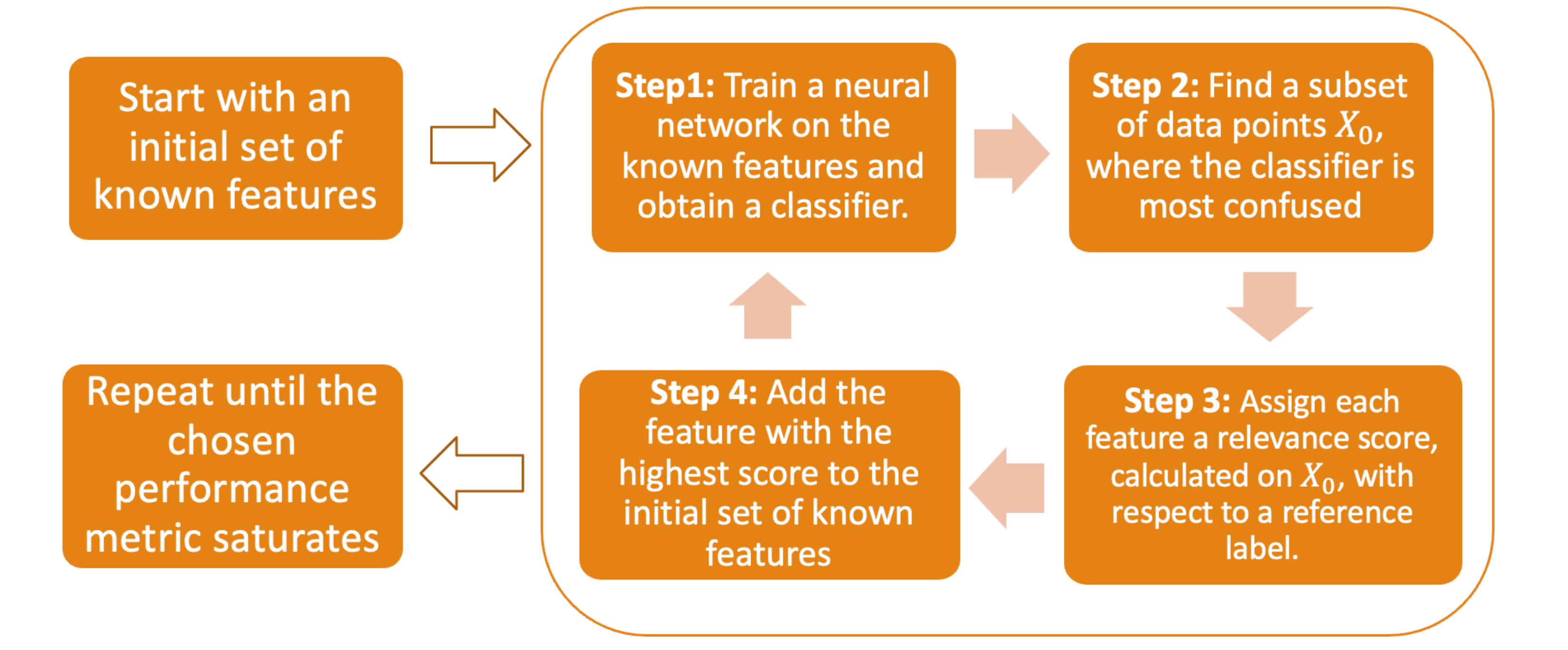}
    \caption{Overview of the proposed forward feature selection algorithm.}
    \label{fig:overview}
\end{figure*}

In this paper, inspired by \cite{ado}, we present a new method for forward feature selection. It is based on the measure of statistical independence called ``distance correlation (DisCo)" \cite{disco,disco_1,disco_2,disco_3}, which was first used  in the HEP literature to decorrelate top taggers against jet mass \cite{DiscoFever}, and was subsequently applied to ABCD background estimation \cite{abcdisco} and anomaly detection \cite{Mikuni:2021nwn}. We use DisCo (instead of ADO) to measure how relevant (statistically dependent) a given set of features is for  the classifier output. We  show that our DisCo-based forward feature selection method outperforms \cite{ado} on both hadronic $W$ tagging and hadronic top tagging, in the sense that it selects features more efficiently, ultimately achieving better performance with fewer features. The upshot is that on top tagging, our method selects as few as 9 EFPs (from the same sample of 7,000+ as \cite{ado}), and training a very compact DNN on these small number of EFPs, we achieve nearly state-of-the-art performance, matching the rejection power of ParticleNet-Lite \cite{particlenet} with only a fraction of parameters.

Importantly, our method does not require a previously-obtained reference classifier, but can also be trained equally well ab initio, using the ``truth labels" (0 for background and 1 for signal). This is unlike the method of \cite{ado}, whose performance suffered when trained on truth labels. Therefore, our DisCo-based forward feature selection method is able to operate in two, conceptually different modes: (1) either  as  an ab-initio feature selector that aims to produce the best-possible classifier given a set of features; or (2) as a feature selector that aims to ``explain" a previously-obtained ``black box" classifier.

Note that the proposed forward or constructive feature selection is very different from \textit{backward}  elimination methods which try to iteratively remove features starting with the full set of features, or feature attribution methods which use Shapley values~\cite{shap_regression,shap_1,shap_2,shap_sampling,deeplift_paper,Shapley,shap,song2016shapley,aas2021explaining,lundberg2019explainable,Sellereite2019}  to assign contributions of each feature to explain the outcome of a pre-trained classifier output. As we will see in the numerical examples, the performance of a classifier trained on the full space of $\approx$~7,000 features is much lower than what a carefully selected set of $\approx 10$ features can achieve, further motivating the forward feature selection strategy.

In the following, we first introduce a strategy for forward feature selection in Section~\ref{section:general_method} and show how DisCo can be used as a scoring function for promising features. Section~\ref{section:top_tagging} next discusses the concrete application to top tagging. 
We show that our method reaches performance equal to much more complex architectures, using only a fraction of features and complexity, even matching LorentzNet~\cite{lorenznet} in ablation studies.
There, we also investigate the leading eight EFPs chosen (as well as their stability under repeated application of our method) and attempt to use them to understand ``what the machine learns". 
We  observe that the same leading six EFPs are found under multiple iterations of our method, indicating their relevance for this task.
Finally, Section~\ref{section_conclusions} provides a discussion of results and further outlook.

\section{Method}\label{section:general_method}

For supervised classification tasks\footnote{In this work, we focus on binary classification as the most widely studied task, but generalisation of the proposed technique to other supervised learning problems is straightforward.}, forward feature selection methods operate on a feature space
\beq
{\mathcal F}=\{f_1,f_2,f_3,\dots,f_N\}
\eeq
We should think of each feature $f_i$ as a pre-determined function (e.g.\ an EFP) that operates on the low-level data $\vec x\in{\Bbb R}^d$ of each event, i.e.\ $f_i=f_i(\vec x)$. Given an already-selected set of $n$ features ${\mathcal F}_n=\{f_{i_1},f_{i_2},\dots,f_{i_n}\}$, the goal of {\it forward} feature selection is to identify the next feature $f_{i_{n+1}}$ which is expected to improve the performance on the classification task the most. 

It is assumed here that the full feature space ${\mathcal F}$ is so large, and the training of the classifier sufficiently expensive, that one cannot just brute force select the next feature by training $N-n$ classifiers on all possible additional features $f_i\notin {\mathcal F}_n$. Therefore, what is needed here is a much cheaper-to-compute {\it relevance score}, that stands in as a proxy for the classifier itself. 

The relevance score takes as input a given set of features, together with a {\it reference label}, evaluated over the dataset. The reference label could be either truth labels, in which case we are performing ab initio forward feature selection in order to produce the highest-performing classifier that we can; or the reference label could be a pre-trained state-of-the-art classifier, in which case we are performing forward feature selection for the purposes of AI explainability (explaining the pre-trained ``black box" classifier). 

In any event, for a set of features, the point is that the relevance score can be obtained much more quickly than training a classifier on the features, and the forward feature selection algorithm can select the feature with the highest score as the next feature.

The 4 steps involved in our feature selection algorithm are illustrated in Fig.~\ref{fig:overview} and explained in the following:

\begin{enumerate}
    \item \textbf{Step 1: Train on known features} 
    
    Train a classifier network on a set of features ${\mathcal F}_n = \{f_{i_1}, f_{i_2}, \dots f_{i_n}\}$ using the full training sample of all events $X_\text{all}$, and obtain the classifier output $y_\text{pred}$
    for all events in $X_\text{all}$.

    For simplicity and best possible performance, we use a dense neural network (details in Appendix~\ref{section:hyperparams}), although any other classification algorithm (e.g.\ XGBoost, logistic regressor) could be used as well.

    \item \textbf{Step 2: Select the confusion set $X_0\subset X_{all}$} 
    
    Instead of calculating the relevance scores using the full dataset, we choose to instead focus on a subset of the full data $X_0\subset X_{all}$ that we call the ``confusion set". These are events where we believe the features in ${\mathcal F}_n$ are least effective in separating signal from background, and where adding a new feature may have the largest impact. To identify this subset, we select all events in a window around $y_{\rm pred}=0.5$, as shown in Fig.~\ref{fig:ypred_window} -- these should be the events where the classifier is most confused about whether it is a signal or a background. We observe that using a confusion set instead of the full dataset improves performance.

\begin{figure}[t]
    \includegraphics[width=\linewidth]{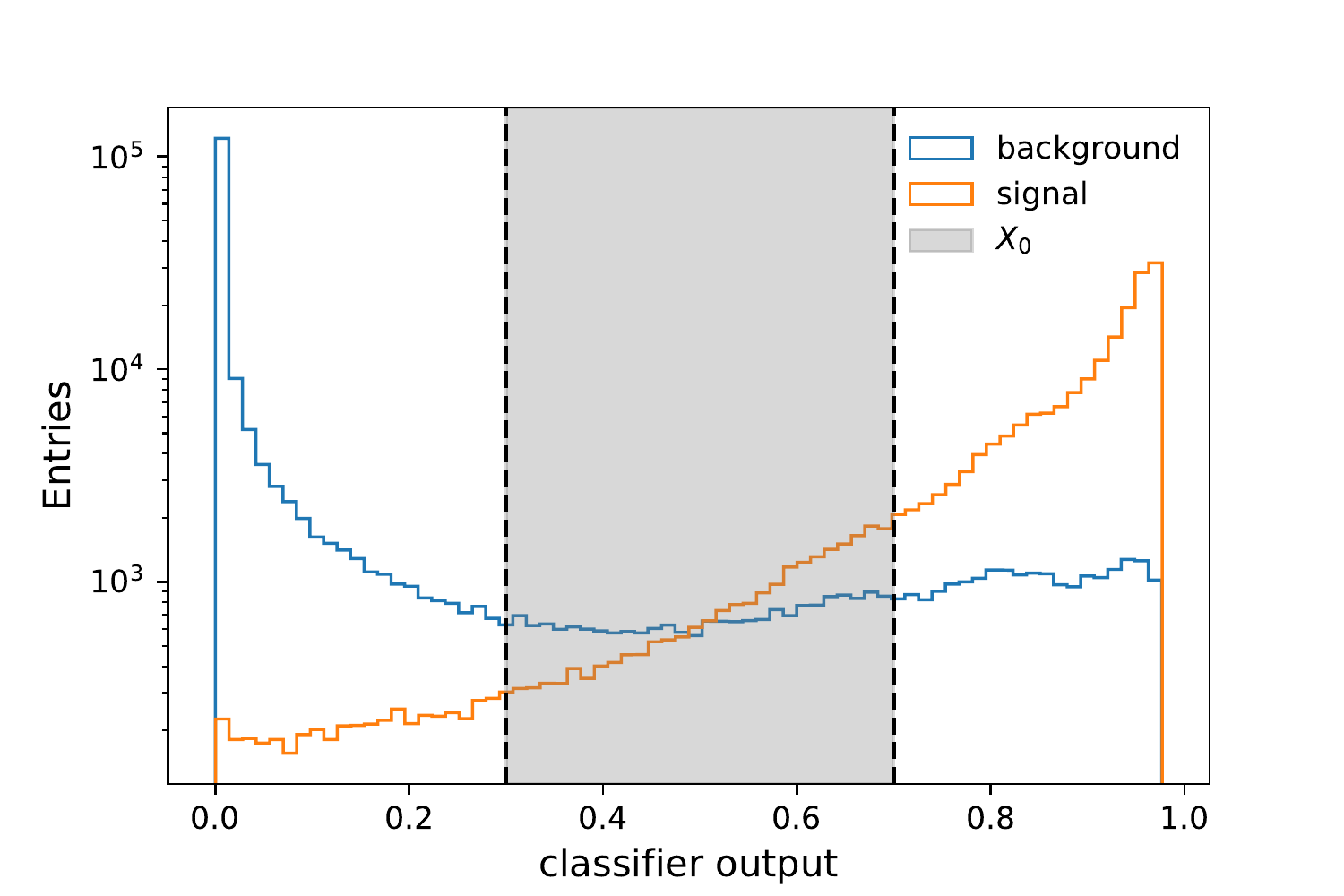}
    \caption{Events in a window around the classifier output value $y_\text{pred}=0.5$ are 
    selected as the confusion set $X_0$ for DisCo-FFS.}
    \label{fig:ypred_window}
\end{figure}

    \item \textbf{Step 3: Assign a relevance score to each feature}
    
    To each feature $f_i$ in the feature space ${\mathcal F}$, we assign a relevance score $s_{f_i}$, which gauges how much the feature will improve classification performance.
    
    The relevance score is calculated using the feature vectors evaluated on the events in the confusion set $X_0$, together with the classifier output of a reference label $y_\text{ref}$ :
    \bea
    {\mathcal X} &=\left\{\Big(f_{i_1}(\vec x),\dots,f_{i_n}(\vec x),f_{i}(\vec x)\Big)\Big|\vec x \in X_0\right\}\\
    {\mathcal Y} &=\{y_\text{ref}(\vec x)|\vec x\in X_0\}
    \eea
    
    The relevance score assigned to each feature $f_i$ is: 
    \begin{equation}
    s_{f_i} = \text{Affine-DisCo}({\mathcal X},{\mathcal Y}).  
    \end{equation}
     
As described in the Introduction, DisCo is short for distance correlation \cite{disco,disco_1,disco_2,disco_3}, a measure of statistical dependence that is zero iff the random vectors ${\mathcal X}$ and ${\mathcal Y}$ are statistically independent, and positive (and $\le 1$) otherwise. Therefore, it is well-suited to judging whether adding $f_i$ to the feature vector $(f_{i_1},\dots f_{i_n})$ produces a stronger correlation with the reference label $y_\text{ref}$ or not. Here we are using the affine-invariant version of DisCo \cite{aff_disco}, which is invariant under arbitrary linear transformations of ${\mathcal X}$ and ${\mathcal Y}$, in order to make it more robust against basis reparametrizations in the EFP space. The multivariate Affine-DisCo calculation is described in more detail in Appendix \ref{section:disco}.

   \item    \textbf{Step 4: Add the feature with best relevance score to the list of known features}
   
     We select the feature with the best score and add it to ${\mathcal F}_n$.
     Then we proceed back to the first step to train a network on the updated set of features ${\mathcal F}_{n+1}$.
     The procedure is stopped when the performance metric saturates and the final set of features is returned.

\end{enumerate}

While the above method explicitly describes our DisCo-based Forward Feature Selection algorithm (DisCo-FFS), the protocol is general enough to accommodate also other iterative feature selection techniques. In Appendix \ref{appendix:ado}, we use the same framework to outline how the Forward Feature Selection from \cite{ado} operates. This is based on Decision Ordering (DO) for the confusion set, and Average Decision Ordering (ADO) for the relevance score, and we will refer to it as DO-ADO-FFS throughout this work.

\section{Application to top-tagging} \label{section:top_tagging}

\subsection{Data set}\label{section:top_data}

\begin{figure*}[t]
    \centering
    \includegraphics[width=\linewidth]{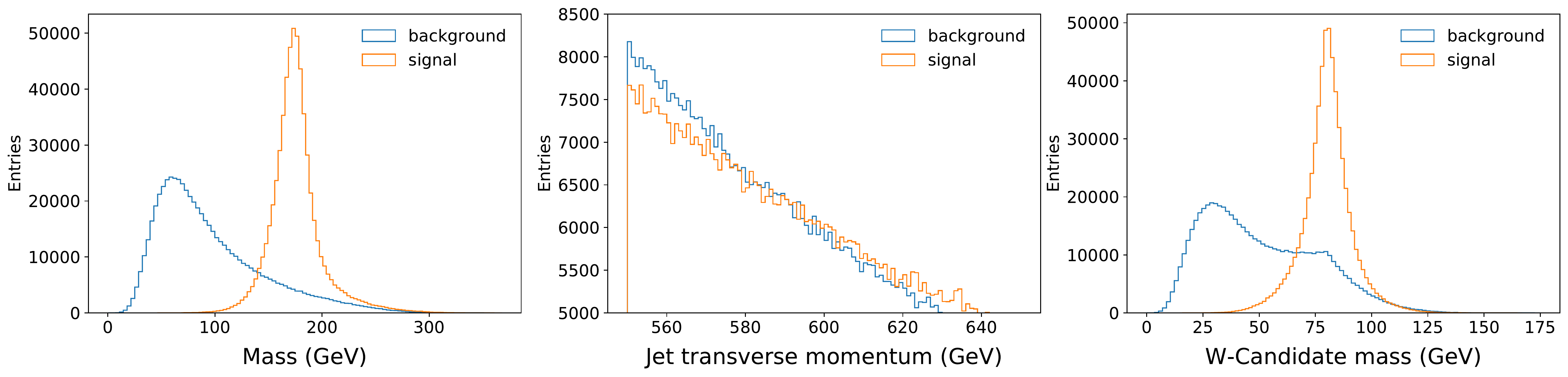}
    \caption{Initial features chosen for top tagging: jet mass $m_J$ (left), jet $p_T$ (center), 
    and mass of the $W$-candidate (right).}
    
    \label{fig:initial_features}
\end{figure*}

We study the performance of the DisCo-Feature Selection algorithms on the top quark tagging landscape data set \cite{landscape,landscape_data}. This data set contains boosted, hadronically-decaying top jets as signal, and QCD (i.e.\ light quark and gluon) jets as background, which are generated using \texttt{Pythia8} \cite{pythia}, with a center-of-mass energy of 14 TeV. Multiple interactions and pile-up are not included in this data set. The detector simulation is done using \texttt{Delphes} \cite{delphes}, with the ATLAS detector card. \texttt{FastJet} \cite{fastjet} is used to create jets using the anti-$k_T$ algorithm \cite{antikt} with $R=0.8$. Only jets in the $p_T$ range $[500,650]$ GeV, and $|\eta_j|<2$, are considered. The data set contains only kinematic information, in the form of energy-momentum four-vectors of all the reconstructed particles in each jet, which are extracted using the Delphes energy-flow algorithm. No additional tracking information or particle information is included. 

The full data set contains 2 million events, with 1 million signal events and 1 million background events. This data is split into $1.2$M events in the training set, 400k in the validation set, and 400k in the test set, each set containing equal number of signal and background events.

\subsection{Feature Space}\label{section:efps}

For top-tagging we start with 
\beq
{\mathcal F}_{initial} ={\mathcal F}_{3} =\{m_J, p_T, m_{W-{\rm candidate}}\}
\eeq
where $m_J$ is the mass of the jet, $p_T$ is the transverse momentum of the jet and $m_{W-{\rm candidate}}$ is the mass of the $W$-candidate in the jet, calculated with a very simple method: we recluster each fat jet using the exclusive $k_T$ algorithm with $R=0.3$ into exactly three subjets. Then we pick the pair of subjets whose invariant mass comes closest to $m_W$. This pair of subjets gives us the $W$-candidate and their mass is $m_{W-{\rm candidate}}$. 
The distributions of the initial features are illustrated in Fig.~\ref{fig:initial_features}.

We then apply feature selection algorithms to a large set of Energy Flow Polynomials (EFPs)\cite{efp}. EFPs are functions of energy fractions and angular separation of jet constituents:
\begin{equation}
z_a^{(\kappa)} = \left( \frac{{p_T}_a}{\sum\limits_{i \in J} {p_T}_i} \right)^{\kappa},\qquad
\theta_{ab}^{(\beta)}=(\Delta \eta_{ab}^2 + \Delta \phi_{ab}^2)^{\beta/2},
\end{equation}
where ${p_T}_a$ is the transverse momentum of the $a$th jet constituent, and the denominator in $z_a$ is summed over all jet constituents in a jet $J$. EFPs have a one-to-one correspondence with a graph $G$:
\beq
\sum_{a \in J} z_a^{(\kappa)} \rightarrow \text{(each node)},\qquad 
\theta_{ab}^{(\beta)} \rightarrow \text{(each edge)}
\eeq
Thus given a graph $G$, with $N$ nodes and edges $(m,\ell) \in G$, the EFP is:
\begin{equation}
\mathrm{EFP}_G^{(\kappa,\beta)}=\sum_{i_1 \in J} \cdots \sum_{i_N \in J} z_{i_1}^{(\kappa)} \cdots z_{i_N}^{(\kappa)} \prod_{(m, \ell) \in G} \theta_{i_m i_{\ell}}^{(\beta)}.
\end{equation}
\noindent

The original EFPs \cite{efp} were introduced as IRC-safe observables, with $\kappa=1$. However in our feature space we are motivated by \cite{ado} to consider other values of $\kappa$ as well. Following \cite{ado},\footnote{With one exception -- we don't include additional features from $d=8$ with $c=4$, as \cite{ado} do in their analysis. These features were initially omitted due to difficulties in their calculation. It was later verified that their inclusion does not significantly alter the performance of DisCo-FFS.} we use Energy Flow Polynomials with all combinations of $d\leq7$, $\beta=[0.5,1,2]$ and $\kappa = [-1,0,0.5,1,2]$, which form a space of 7,320 unique features.

\subsection{Results}\label{section:top_disco_fs}

\subsubsection{Ab initio feature selection using truth labels}

First, we consider the ab initio feature selection task, using the truth labels to guide the algorithms so as to yield the best-possible classifier. 

We apply the truth-guided DisCo-FFS and DO-ADO-FFS\footnote{We note that in \cite{ado}, the DO with truth-labels was referred to as TO (for ``truth-ordering") and it was pointed out that ADO with truth-labels reduces to the usual AUC metric.} to the training and validation set, and use the test set only for evaluating the performance. (Network architectures and hyperparameters used in this section are described in Appendix \ref{section:hyperparams}.) 
The performance metric choosen for top-tagging is $R_{30}$ (the QCD rejection factor at 30\% top-tagging efficiency). 
It allows a better separation of different methods as area under curve (AUC) saturates and is more indicative of the performance
at a potential working point.

As shown in Fig.~\ref{fig:performance}, the $R_{30}$ value increases as more features are added using the two feature selection methods.
This shows that both DisCo-FFS and DO-ADO-FFS are selecting useful features. After 9 features the performance of the features added using the DisCo method saturates with $R_{30}\approx 1250$. We also see that our proposed method outperforms DO-ADO-FFS and achieves a higher $R_{30}$ at each step.

Any worthwhile feature selection algorithm should do better than randomly selecting features.  To test this, we randomly select each number of features 10 times, and use the average and standard deviation of the $R_{30}$ as our ``random baseline" shown in Fig.~\ref{fig:performance}. Interestingly we see that the randomly selecting EFPs can also give better performance, as we add more and more features, but not as good as the FFS methods.

\subsubsection{Feature selection using pre-trained classifier}

Next we turn to feature selection using a pre-trained classifier (so-called ``black-box guiding" in \cite{ado}). For the pre-trained classifier, we use the state-of-the-art \texttt{LorentzNet} tagger \cite{lorenznet}. 

\begin{figure*}
 \centering 
  \includegraphics[width=0.75\linewidth]{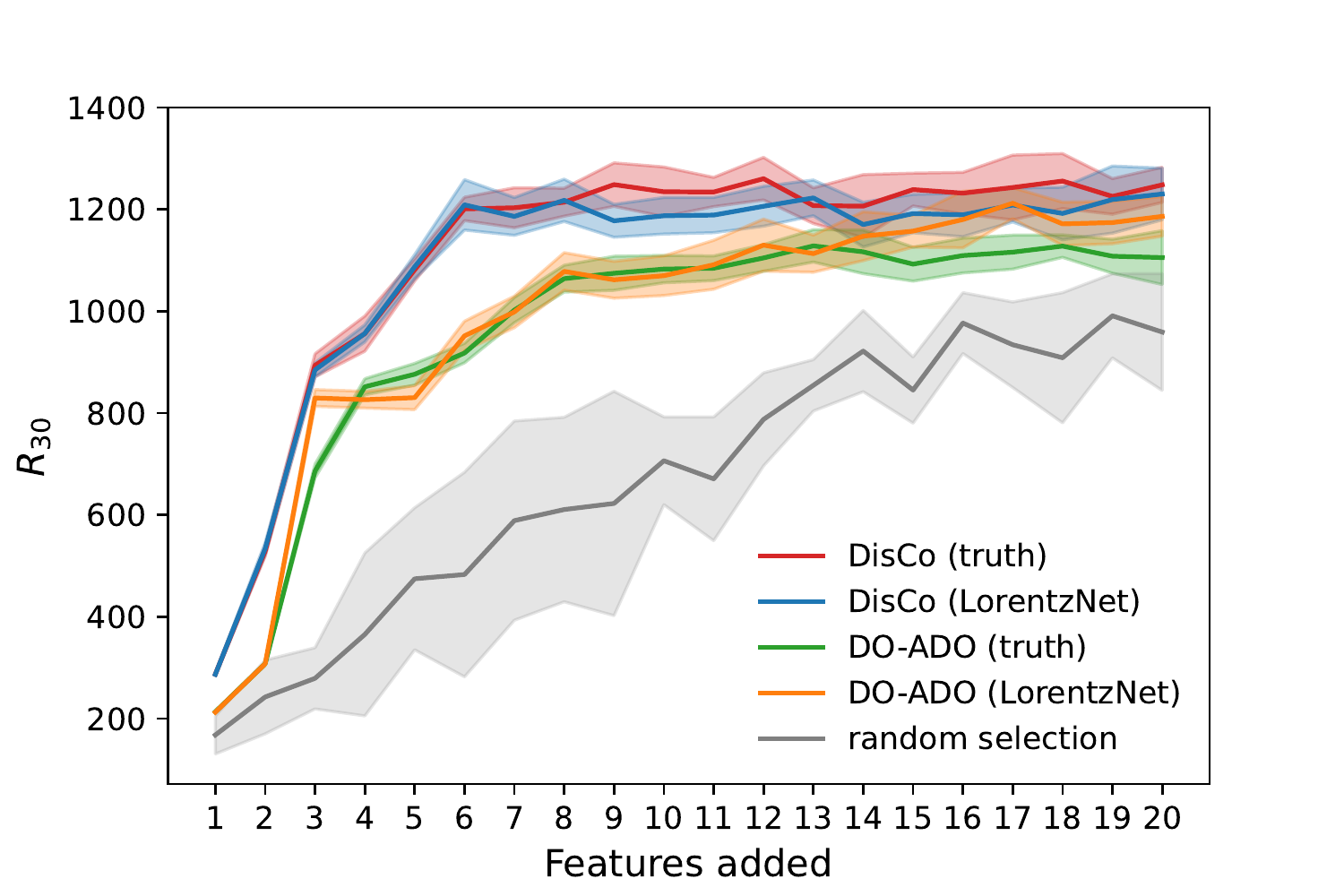}
\caption{Performance comparison between
DisCo-FFS and DO-ADO-FFS methods, truth-guided and \texttt{LorentzNet}-guided. Shown in gray is also the random selection baseline. The shaded bands around each curve come from training the NN classifier ten times on the same set of features (similar to \cite{landscape}). Overall, DisCo-FFS seems to select more relevant features than DO-ADO-FFS, resulting in a higher-performing classifier at every step. Interestingly, 
while DO-ADO-FFS with truth labels actually performs {\it worse} than with \texttt{LorentzNet} (a phenomenon also observed in \cite{ado}), no degradation in performance is observed for DisCo-FFS with truth labels.}
\label{fig:performance}
\end{figure*}

We see in Fig.~\ref{fig:performance} that DO-ADO-FFS with \texttt{LorentzNet} actually performs slightly {\it better} than DO-ADO-FFS with truth labels. This somewhat counterintuitive result was also observed by \cite{ado} in the context of boosted $W$-tagging, and we confirm it here.  As explained there, the confusion set of the DO-ADO method consists of signal-background pairs which are incorrectly ordered by the classifier trained at every step (called $y_{\rm pred}$ in Sec.~\ref{section:general_method}), with respect to the reference labels. When using truth labels for the latter, the confusion set can be significantly contaminated by signal-background pairs which may never be ordered properly, even by the ideal Neyman-Pearson classifier. This can in turn distort the ADO score which is calculated on the confusion set. This explains why the \texttt{LorentzNet}-guided DO-ADO-FFS performs better than the truth-guided DO-ADO-FFS.

Meanwhile, we see from Fig.~\ref{fig:performance} that there is no significant difference in performance between truth-guided and \texttt{LorentzNet}-guided DisCo-FFS. This is perhaps the more expected and intuitive result. We believe the reason DisCo-FFS does not suffer from the degradation in performance when using truth labels can be understood by the fact that our confusion set is determined solely using the classifier trained at every step, and does not involve the reference labels at all. Also, our confusion set is determined on background and signal jets separately. Therefore, the issue of the forever-incorrectly-ordered signal-background pairs never even arises here. It would be interesting to test this explanation further, for example by combining these different ways of choosing the confusion set (DO or $y_{pred}$) with different relevance scores (ADO or DisCo). We reserve this for future work.

In any case, we conclude that, unlike DO-ADO-FFS, DisCo-FFS does not seem suffer in performance when using truth labels instead of a state-of-the-art pre-trained tagger. This means that DisCo-FFS should be a suitable method for both ab initio feature selection and for explaining black-box taggers.

\subsection{Comparison with other taggers}

\begin{table}[t]
    \centering
    \begin{tabular}{|l|c|c|l|}
        \hline Taggers & AUC & $R_{30}$ & Param \\
          \hline
 Linear 1k EFPs \cite{efp}  & 0.980  & 384 & 1 $\mathrm{k}$ \\   N-sub 6 \cite{nsub_landscape} & 0.979 & 792 $\pm$ 18 & 57 $\mathrm{k}$ \\
   N-sub 8 \cite{nsub_landscape} & 0.981 & 867 $\pm$ 15 & 58 $\mathrm{k}$  \\ \hline   ParticleNet \cite{particlenet} & 0.986 & 1615 $\pm$ 93 & 366 $\mathrm{k}$ \\   ParticleNet-Lite \cite{particlenet} & 0.984 & 1262 $\pm$ 49 & 26 $\mathrm{k}$ \\   LorentzNet \cite{lorenznet} & 0.987 & 2195 $\pm$ 173 & 224 $\mathrm{k}$ \\   ParT \cite{part} & 0.986 & 1602 $\pm$ 81 & 2.14 $\mathrm{M}$ \\   PELICAN \cite{pelican} & 0.987 & 2289 $\pm$ 204 & 45 $\mathrm{k}$ \\ \hline  DNN 7k EFPs & 0.980 & 844 & 237 $\mathrm{k}$  \\
  DO-ADO (\texttt{LorentzNet}) & 0.982 & 1212 $\pm$ 30 & 1.7 $\mathrm{k}$  \\  \textbf{DisCo-FFS (truth)} & 0.982 & 1249 $\pm 43$ & 1.4 $\mathrm{k}$  \\ 
\hline
    \end{tabular}
    \caption{AUC and $R_{30}$ comparison of different taggers on the dataset from \cite{landscape}. The $R_{30}$ values of DisCo-FFS and ADO-FFS are the average $R_{30}$'s of 10 classifier trainings, and the $R_{30}$ of DNN on 7k EFPs is calculated over a single run. The performance for DisCo-FFS is after 9 EFPs, whereas the performance reported for DO-ADO is after 17 EFPs. }
    \label{table:performance}
\end{table}

\begin{figure*}
    \centering
    \includegraphics[width=0.75\linewidth]{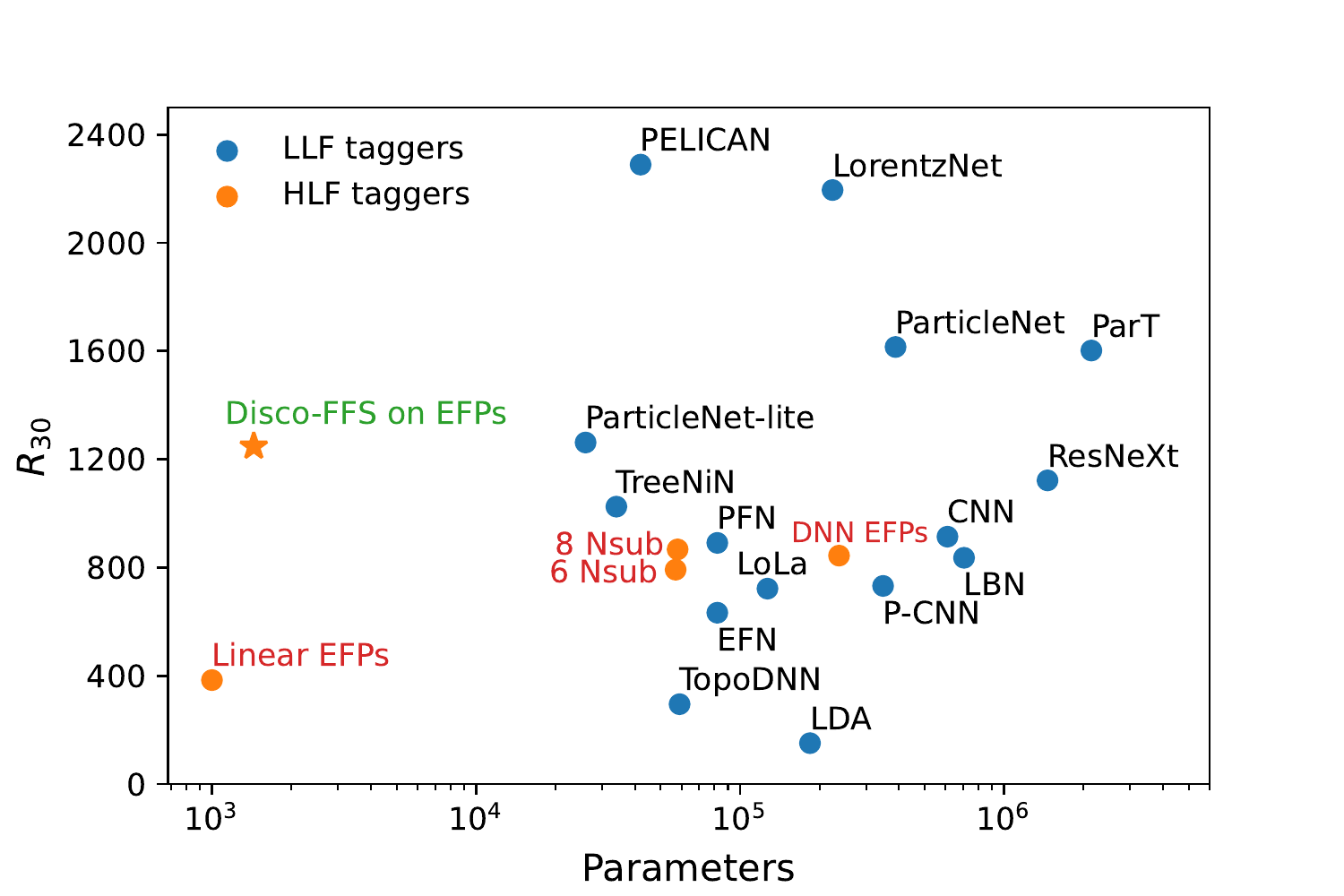}    \caption{$R_{30}$ vs.\ number of parameters of the model, for many different approaches to top-tagging. \texttt{LorentzNet}\cite{lorenznet}, \texttt{PaticleNet} \cite{particlenet}, \texttt{ParT} \cite{part}, and \texttt{PELICAN} \cite{pelican} are the some of the recent taggers with very good performances. ``DisCo-FFS on EFPs" 
corresponds to the simple DNN trained on the first nine EFPs selected by DisCo-FFS, while ``DNN EFPs" is our DNN trained on all the 7k EFPs. The remaining taggers are taken from \cite{landscape}. We see that the nine EFPs selected using Disco-FFS have a very competitive performance, especially given the number of parameters.}
    \label{fig:comparison_taggers}
\end{figure*}

The top-tagging comparison study \cite{landscape} includes two methods which use high-level features as inputs for top-tagging: one used a NN with multi-body $N$-subjettiness as input features \cite{nsub_landscape,dutta_landscape}, and the other uses a Linear Classification (with Fischer's Linear Discriminant) on EFPs. All other taggers are based on low-level jet information. 
The proposed DisCo-FFS selection strategy based on 9 EFPs and 3 initial features outperforms all methods in the published study \cite{landscape}. However, it falls short in performance to even more state-of-the-art taggers that were published after \cite{landscape}: \texttt{ParticleNet} \cite{particlenet}, \texttt{LorentzNet} \cite{lorenznet}, the \texttt{ParT} (particle transformer net) tagger \cite{part}, and \texttt{PELICAN} \cite{pelican}. Nevertheless, our tagger is able to achieve a very competitive performance with only 1440 parameters as shown in Table \ref{table:performance} and Fig.~\ref{fig:comparison_taggers}.

We also compare our performance to that of a network (architecture described in Appendix \ref{section:hyperparams}) that was trained on all 7k EFPs. As shown in Table~\ref{table:performance}, this network is only able to obtain a performance of $R_{30}=844$. This is significantly worse than the performance using the small subset of EFPs selected by DisCo-FFS.  Clearly, the use of uninformative features in the training deteriorates the performance of the network. In principle,  it should be possible to optimize the hyper-parameters to recover the lost performance, but this is not so straightforward in practice, given the amount of time and resources it takes to train a network on all 7k EFPs.\footnote{This is also why the $R_{30}$ quoted here does not come with an error bar from multiple retrainings --  a single training was already prohibitively time consuming for us.} This emphasizes the need of doing feature selection.

As a further aside, this result also indicates why another popular feature selection method, which is based on assigning feature attributions using Shapley values, is not suitable here. Shapley values assume the existence of a high-performing classifier trained on a set of features, and then ranks those features in terms of their estimated contributions to the classifier outputs. In fact, the original Shapley values \cite{Shapley,shap_1,shap_2} are very much ill-suited to the problem at hand -- their computational complexity grows exponentially with the number of features, so in practice can never be computed for more than $\sim 10$ features. Also the features are assumed to be uncorrelated, for the computation of Shapley values.  With 7k highly correlated features, this is clearly not the right approach. Later approaches such as SHAP \cite{shap} attempt to overcome the computational complexity issue by approximating the Shapley values in various ways. SHAP also used (approximate) Shapley values to unify different feature attribution methods \cite{lime,deeplift_paper,shap_regression,shap_sampling}. But generally all these works still assume independence of the features. This is an area of active research and it is possible a Shapley-inspired approach will work well on this problem in the future. Suffice to say that in our experiments (based on Deep SHAP \cite{shap,deeplift_paper} and the sub-par DNN trained on 7k EFPs), we obtained results that were only marginally better than random selection.

\subsection{Ablation studies}

\begin{figure}
    \centering
    \includegraphics[width=\linewidth]{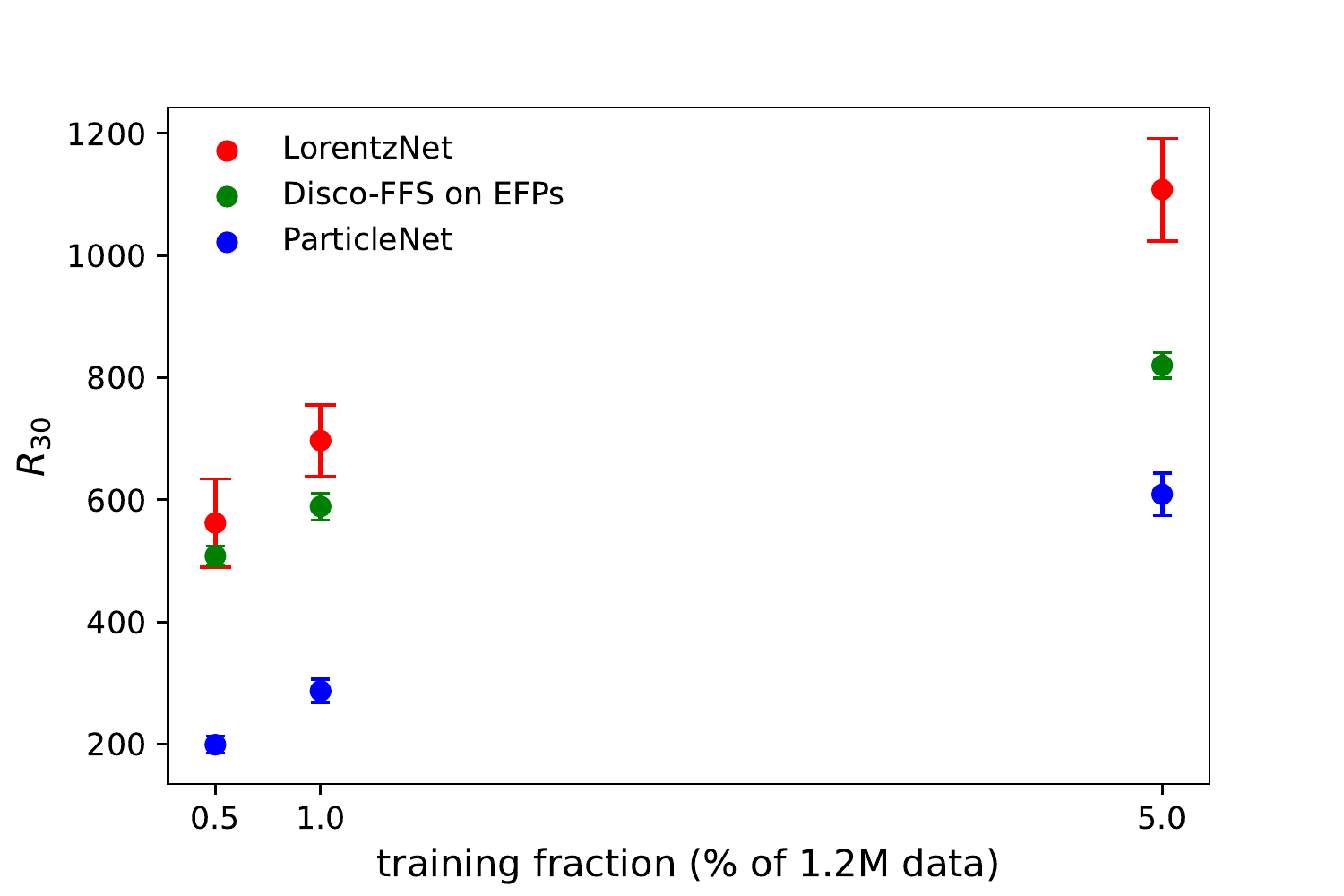}
    \caption{Performance of training on 0.5\%, 1\% and 5\% of the training data. The EFPs selected using DisCo outperform \texttt{ParticleNet}, and match up to the performance of \texttt{LorentzNet} \cite{lorenznet} at 0.5\% of the total training data.}
    \label{fig:sample_efficiency}
\end{figure}

To showcase another important benefit of feature selection, we compare the performance of the features we obtained using DisCo-FFS to \texttt{ParticleNet} and \texttt{LorentzNet}, on smaller training datasets. We take the set of features obtained in section \ref{section:top_disco_fs} and train the same neural network with same hyper-parameters on $5\%$, $1\%$ and $0.5\%$ of the same training data. While both \texttt{LorentzNet} and \texttt{ParticleNet} had a superior performance for the full training dataset, our set of features outperforms \texttt{ParticleNet} at lower training fractions, and more-or-less matches \texttt{LorentzNet} at $0.5\%$ and $1\%$ of the training dataset, as shown in Fig.~\ref{fig:sample_efficiency}.

\begin{figure*}[t]
\centering
    \includegraphics[width=0.75\linewidth]{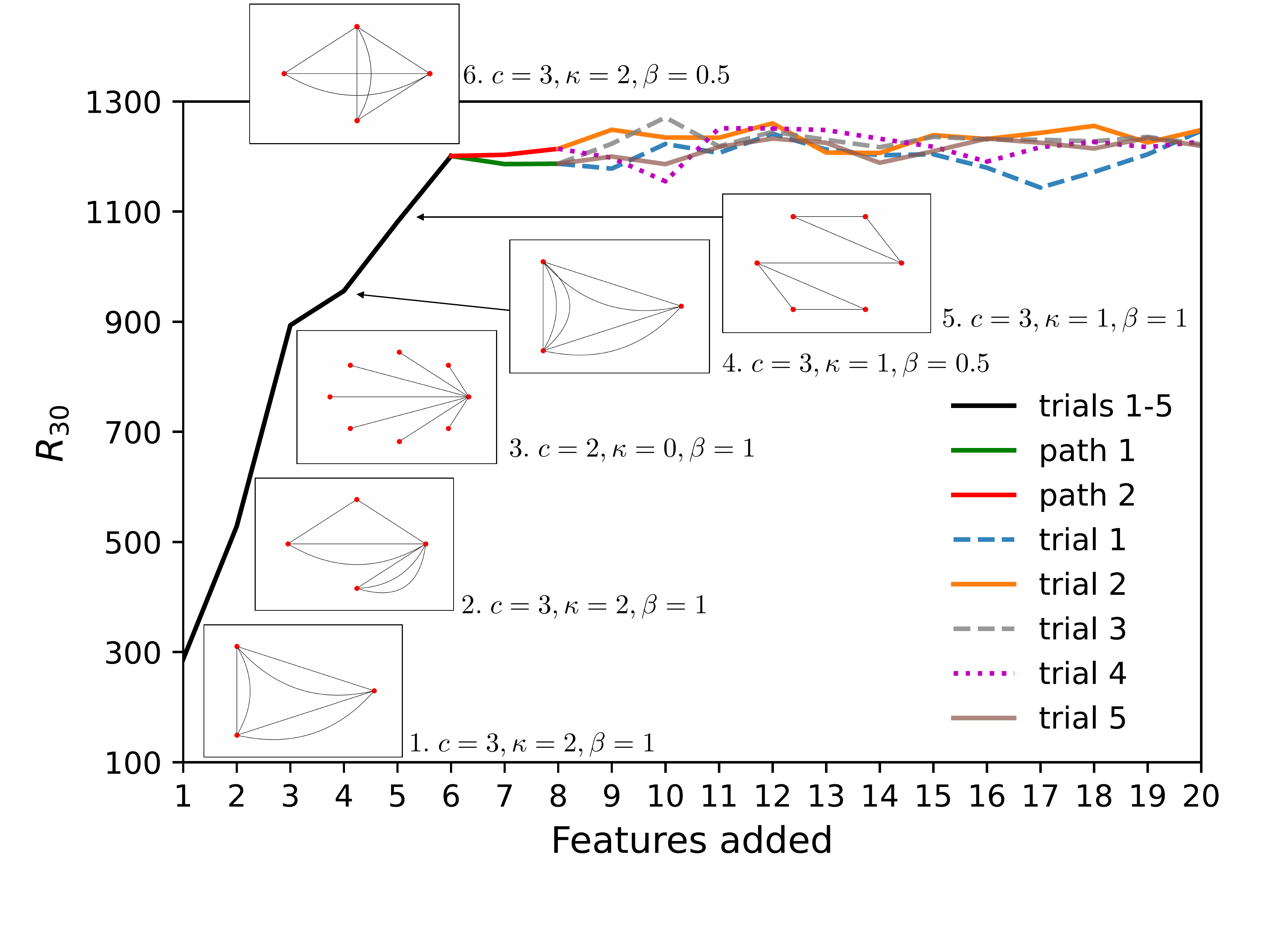}
\caption{Performance vs.\ iteration for 5 trials of DisCo-FFS (performance is the mean $R_{30}$ of 10 trainings). We see that the feature selection is deterministic for the first six EFPs selected (superimposed), and there is a corresponding sharp rise in $R_{30}$. Then this is followed by 2 paths (marked path 1 and path 2) in the $7^\text{th}$ and $8^\text{th}$ iterations.  After that, DisCo-FFS finds different sets of features to achieve similar performance.  }
    \label{fig:unique_features}

\end{figure*}

\subsection{Robustness of the feature selection}

It is interesting to ask whether the DisCo-FFS algorithm selects the same features every time. This is not a priori guaranteed, because there is some stochasticity to the algorithm, coming from the training of the NN classifier at every step (which in turn determines the confusion set on which the relevance score is calculated). 

Shown in Fig.~\ref{fig:unique_features} is the $R_{30}$ vs number of features selected, after running the DisCo-FFS algorithm five independent times. We see that DisCo-FFS  repeatedly chooses the same first six EFPs. After that, the features selected start to diverge from fully deterministic, at first only slowly (there appear to be two possibilities for the pairs of EFPs selected in the 7th and 8th iterations), and then quickly from the 9th EFP onwards (on the 9th EFP, the five trials selected five different EFPs).

This is broadly consistent with Fig.~\ref{fig:performance}. There we see the $R_{30}$ shooting up rapidly during the first six EFPs, indicating that they provide a lot of classification power, and should produce a strong signal for the relevance score in the DisCo-FFS selection procedure. Then the $R_{30}$ plateaus but does rise a little bit, from six EFPs to nine EFPs. This is consistent with a much weaker signal coming from the relevance score and more possibility for randomness. Finally, after nine EFPs, the $R_{30}$ no longer rises and instead fluctuates around 1250. This is consistent with the remaining EFPs being selected randomly and not providing any real signal to the relevance score.

\begin{table}[t]
\centering
  \begin{tabular}{| c | p{1.2cm} | >{\centering}p{0.5cm} |  >{\centering}p{0.5cm} |  >{\centering}p{0.5cm} |   c |}
    \hline
    
         Iter & Feature & c & $\kappa$ & $\beta$  & $R_{30}$\\ 
    \hline
        1 & \includegraphics[width=\linewidth, height=10mm]{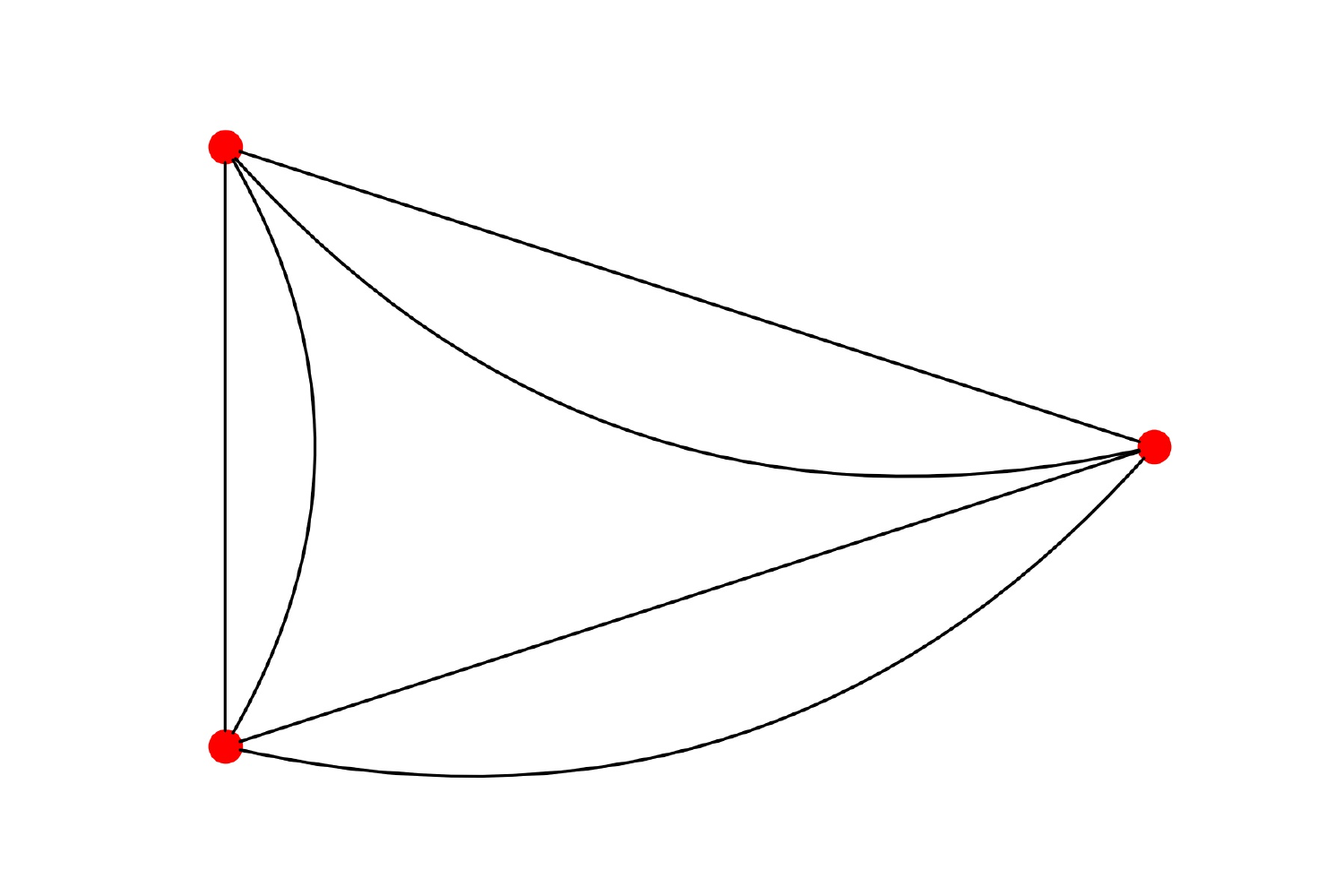} & 3 & 2 & 1 & 287 $\pm$ 3 \\ 
    \hline
        2 & \includegraphics[width=\linewidth, height=10mm]{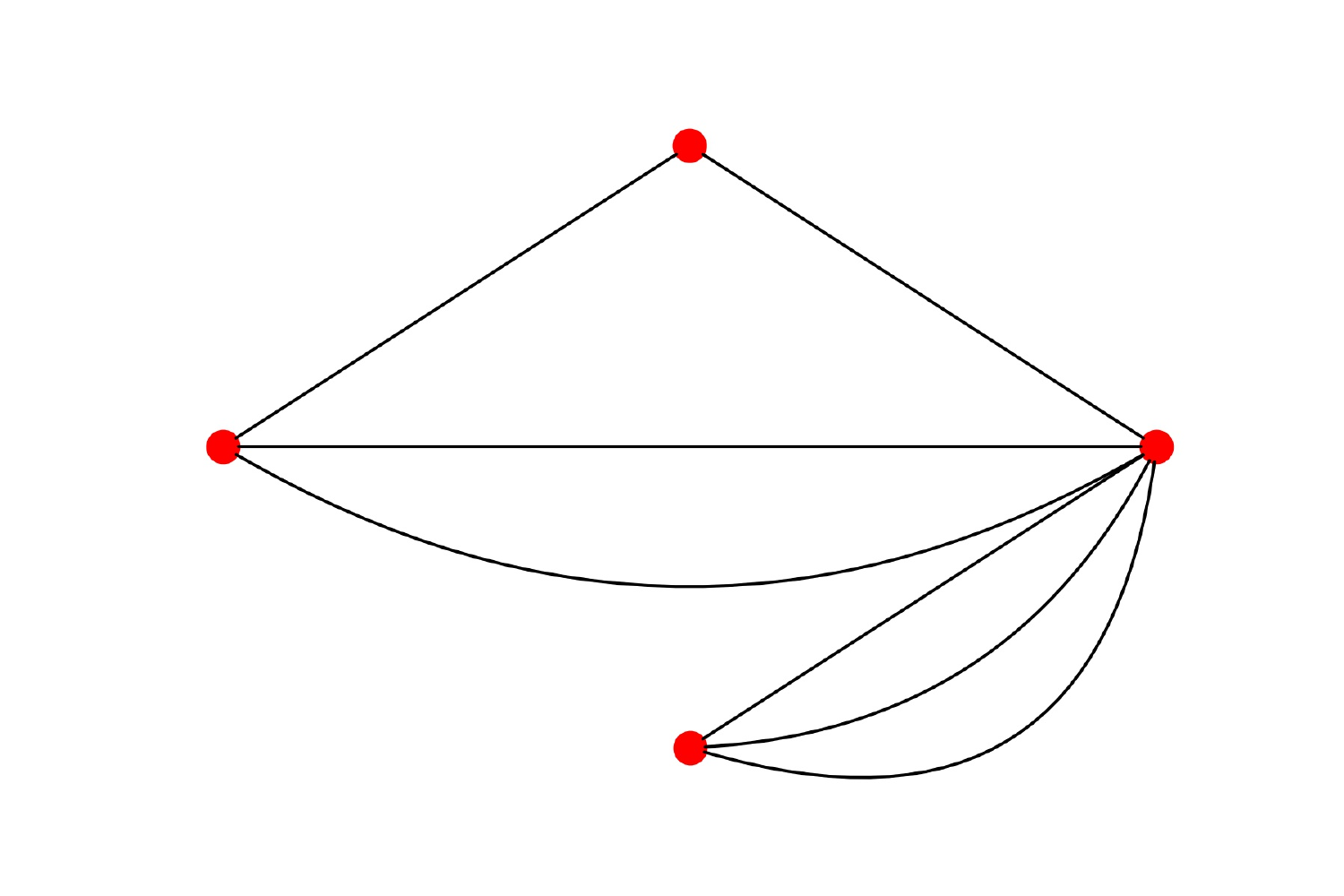} & 3 & 2 & 1 & 529 $\pm$ 10\\ 
    \hline
        3 & \includegraphics[width=\linewidth, height=10mm]{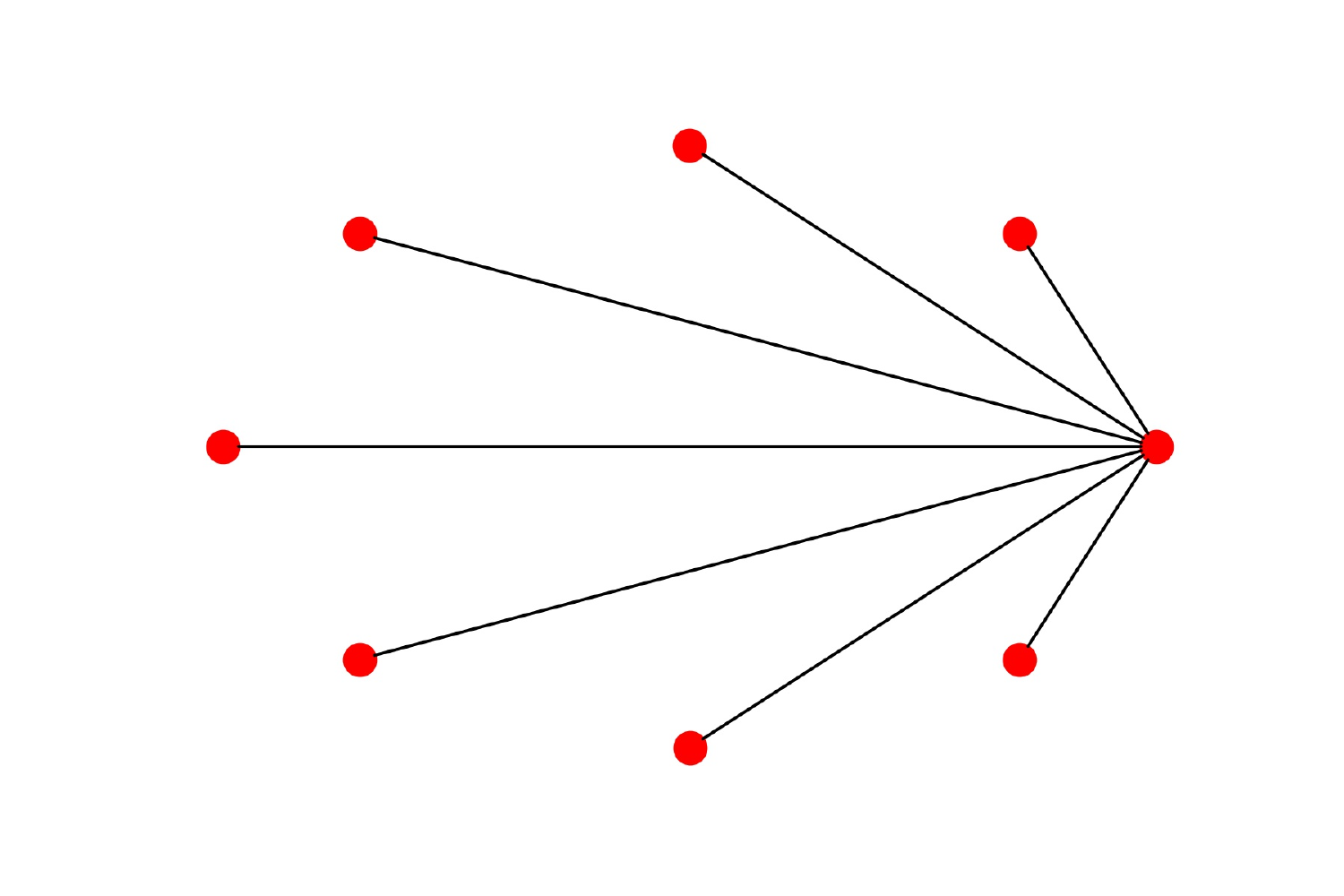} & 2 & 0 & 1 & 894 $\pm$ 23 \\ 
    \hline    
        4 & \includegraphics[width=\linewidth, height=10mm]{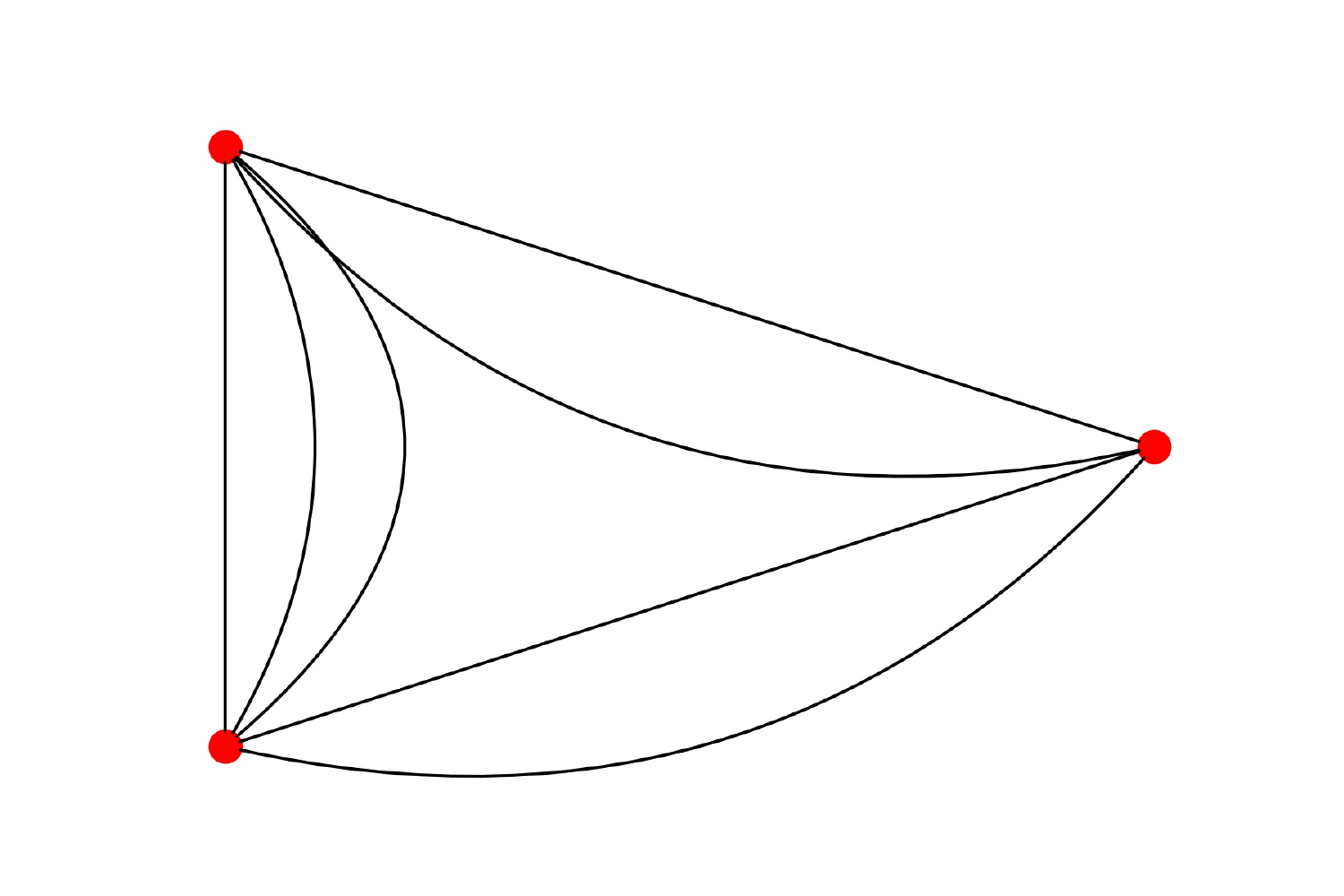} & 3 & 1 & 0.5 & 956 $\pm$ 35\\ 
     \hline   
        5 & \includegraphics[width=\linewidth, height=10mm]{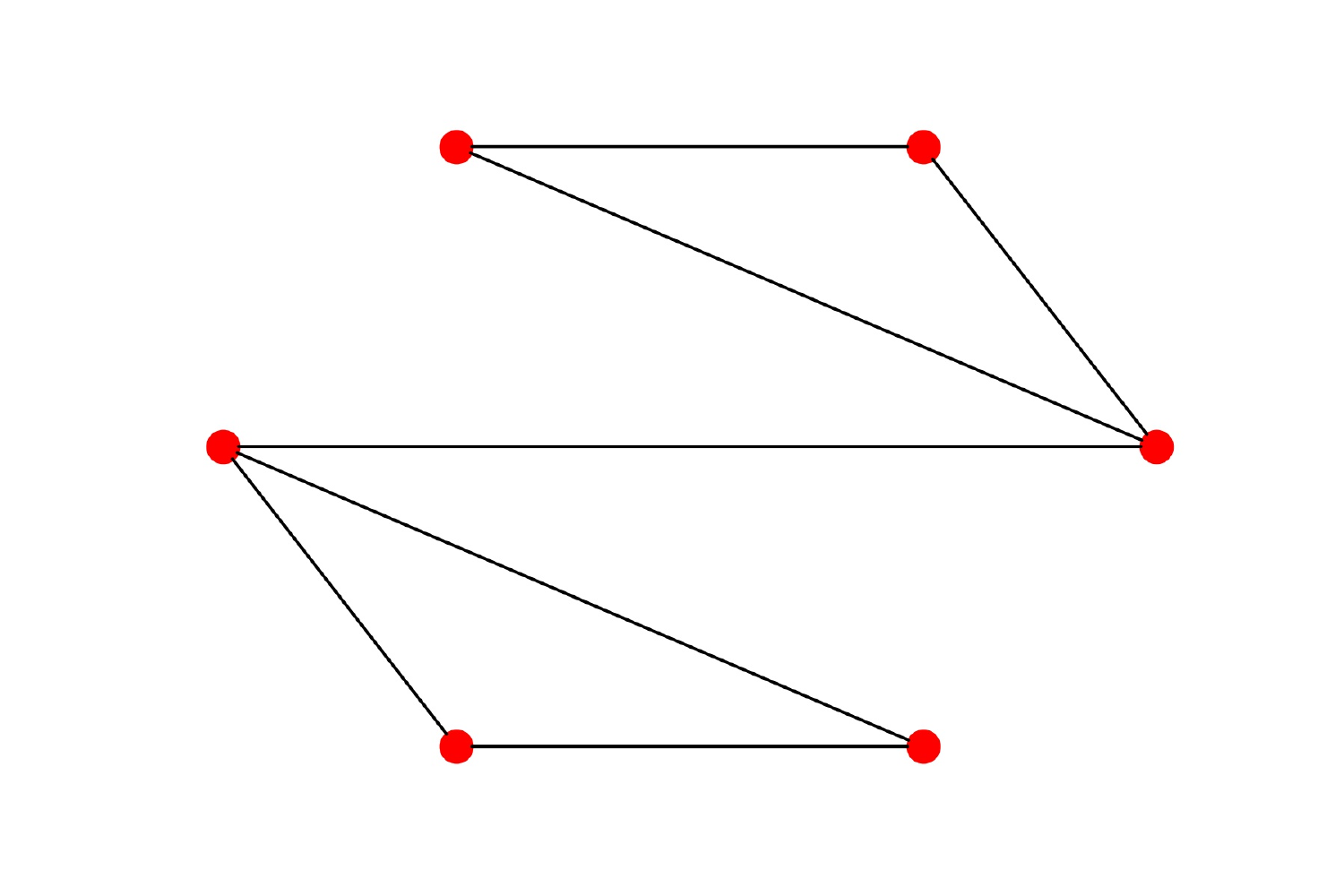} & 3 & 1 & 1 & 1081 $\pm$ 22\\ 
     \hline   
        6 & \includegraphics[width=\linewidth, height=10mm]{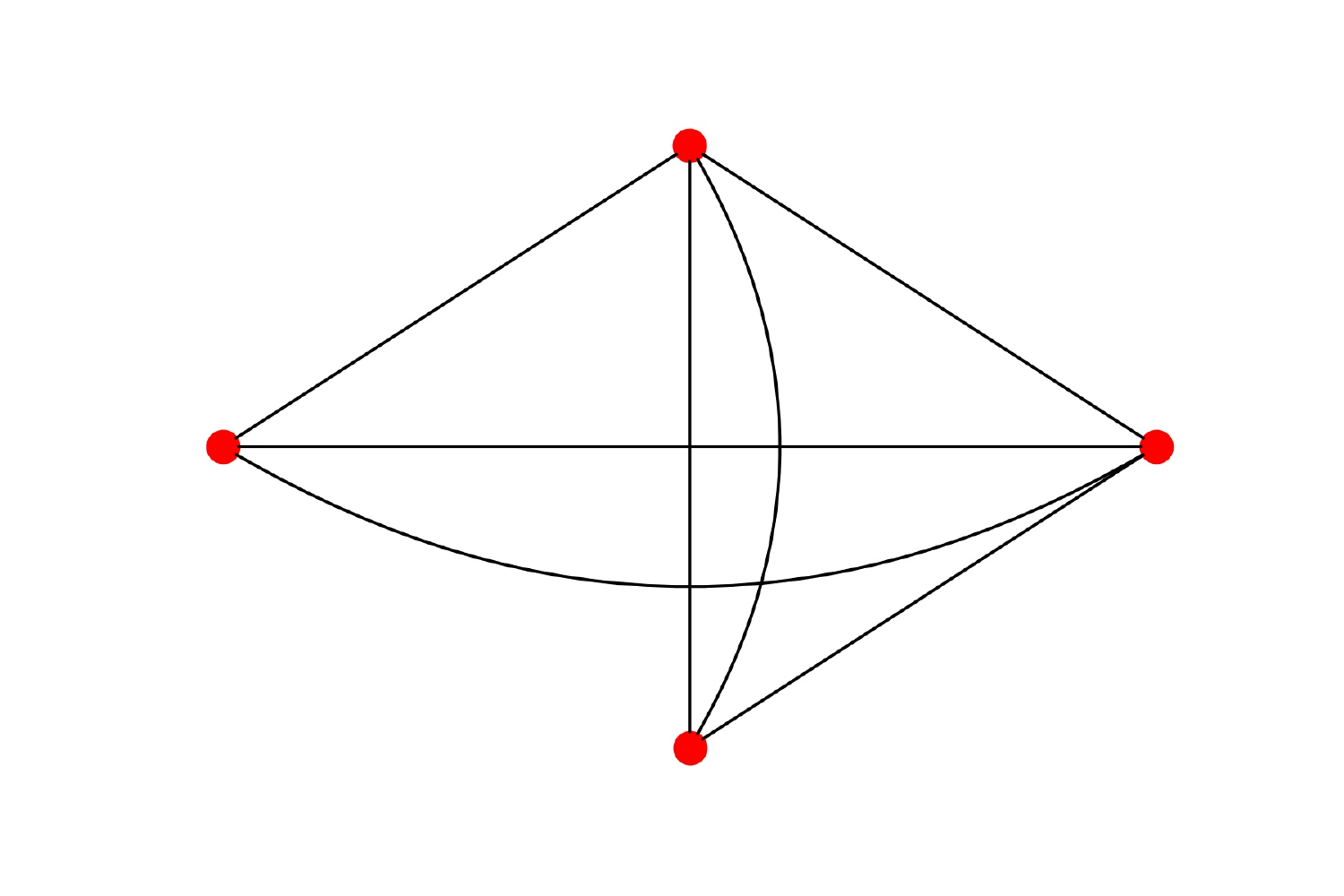} & 3 & 2 & 0.5 & 1201 $\pm$ 23\\ 
     \hline

  \end{tabular}
     \caption{The EFPs selected by Disco-FFS in the first 6 iterations}
     \label{tbl:efps}
      \end{table}

\begin{table}[t]

       \textbf{Path 1: }
        \begin{tabular}{| c | m{1.2cm} |>{\centering}p{0.4cm} |>{\centering}p{0.4cm} | c |}
        \hline
         Iter & Feature & c & $\kappa$ & $\beta$ \\
         \hline
        7 & \includegraphics[width=\linewidth, height=10mm]{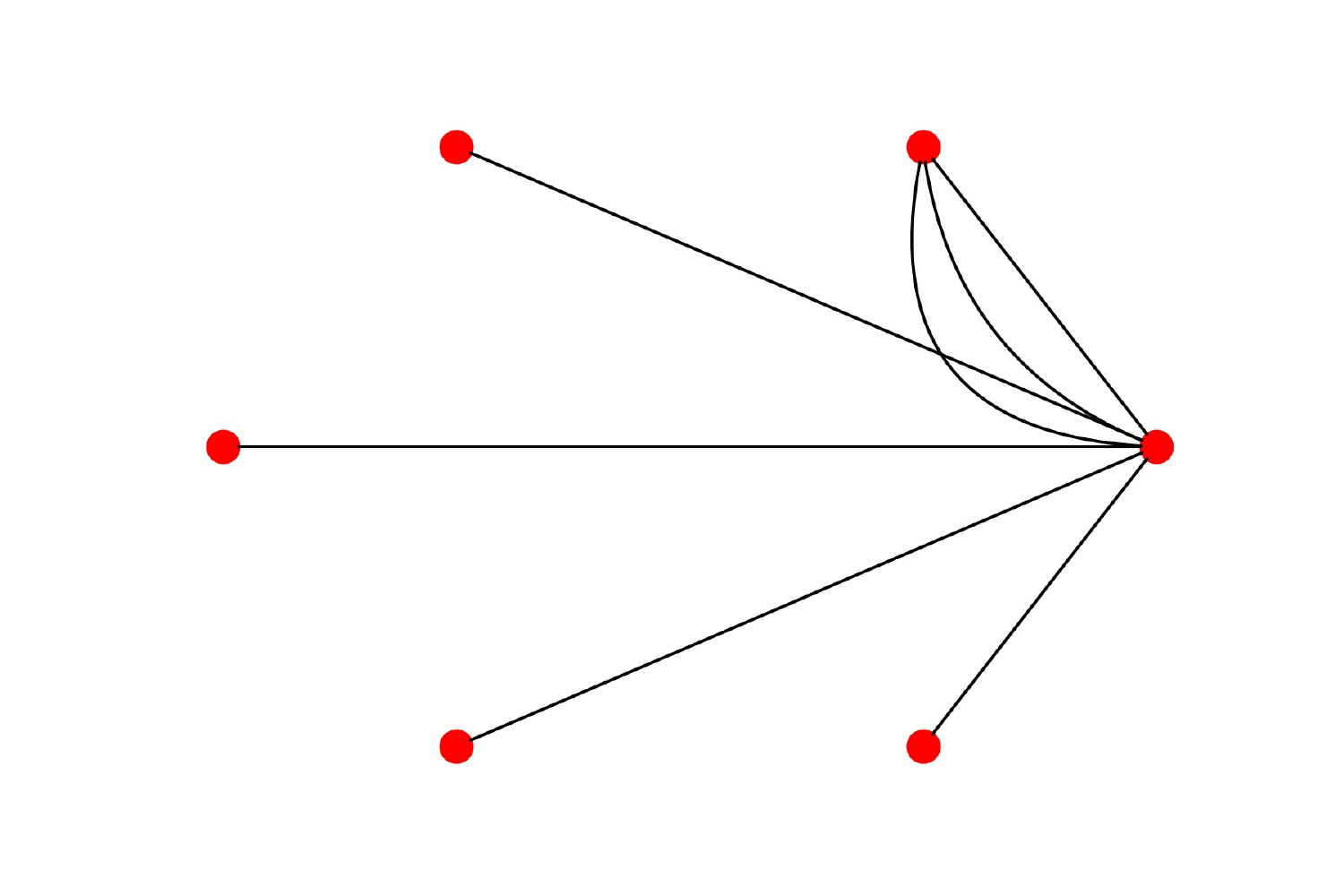} & 2 & 0 & 0.5 \\ 
     \hline   
        8 & \includegraphics[width=\linewidth, height=10mm]{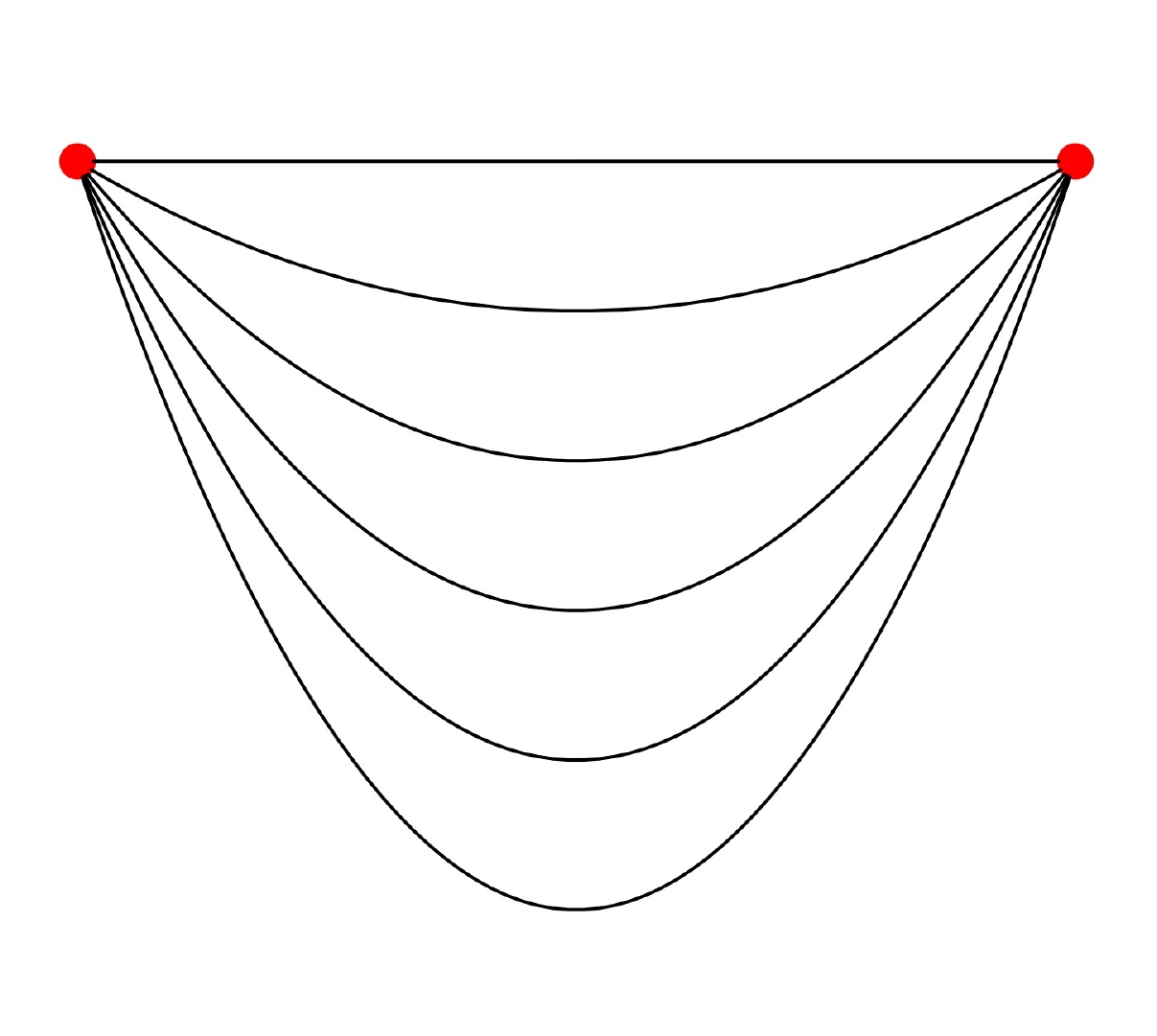} & 2 & 2 & 2 \\  
        \hline
        \end{tabular}

\vspace{10pt}
      \centering

       \textbf{Path 2: }
        \begin{tabular}{| c | m{1.2cm} |>{\centering}p{0.4cm} |  c | c |}
         \hline 
          Iter & Feature & c & $\kappa$ & $\beta$ \\
          \hline
        7 & \includegraphics[width=\linewidth, height=10mm]{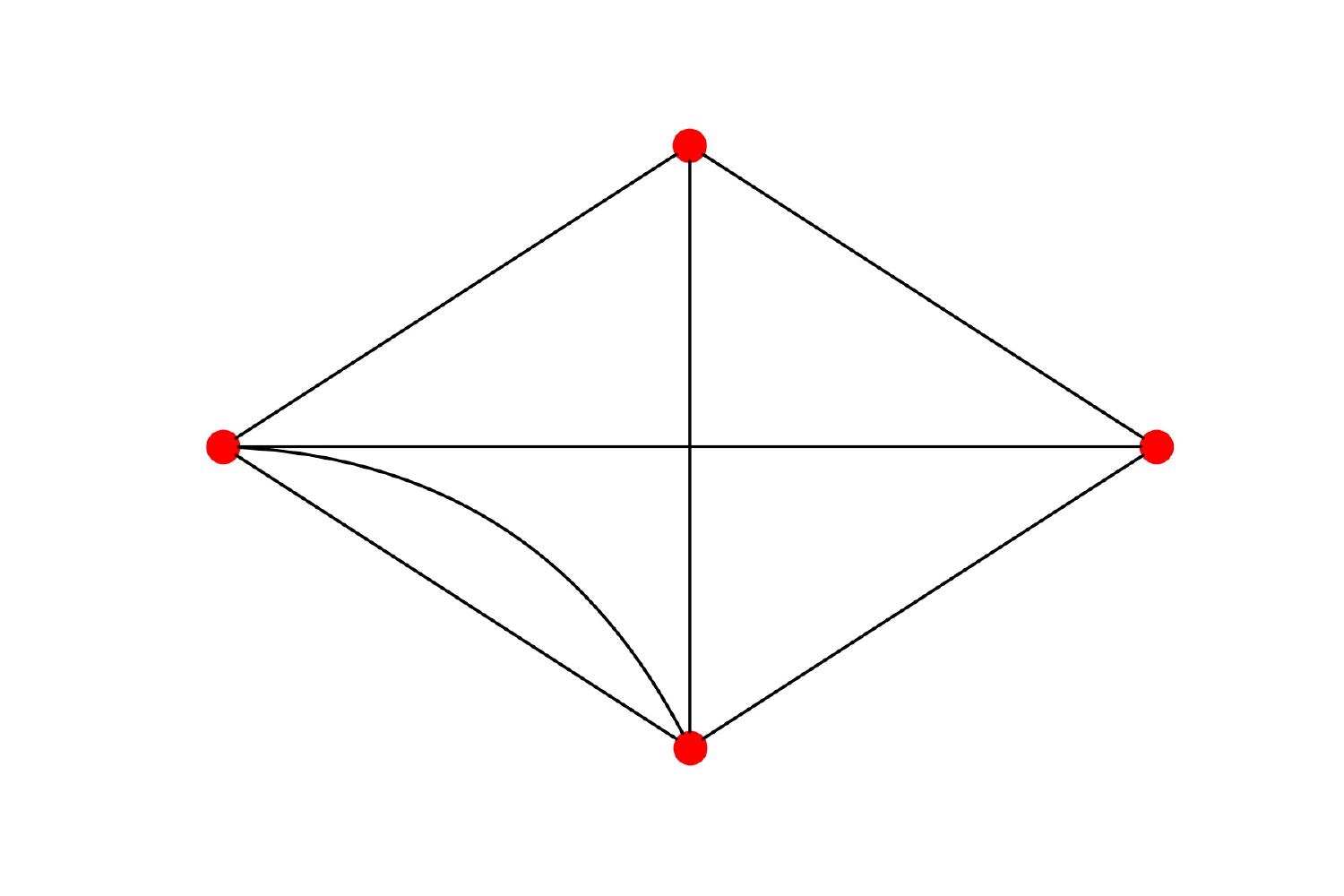} & 4 & 0.5 & 0.5 \\ 
     \hline   
        8 & \includegraphics[width=\linewidth, height=10mm]{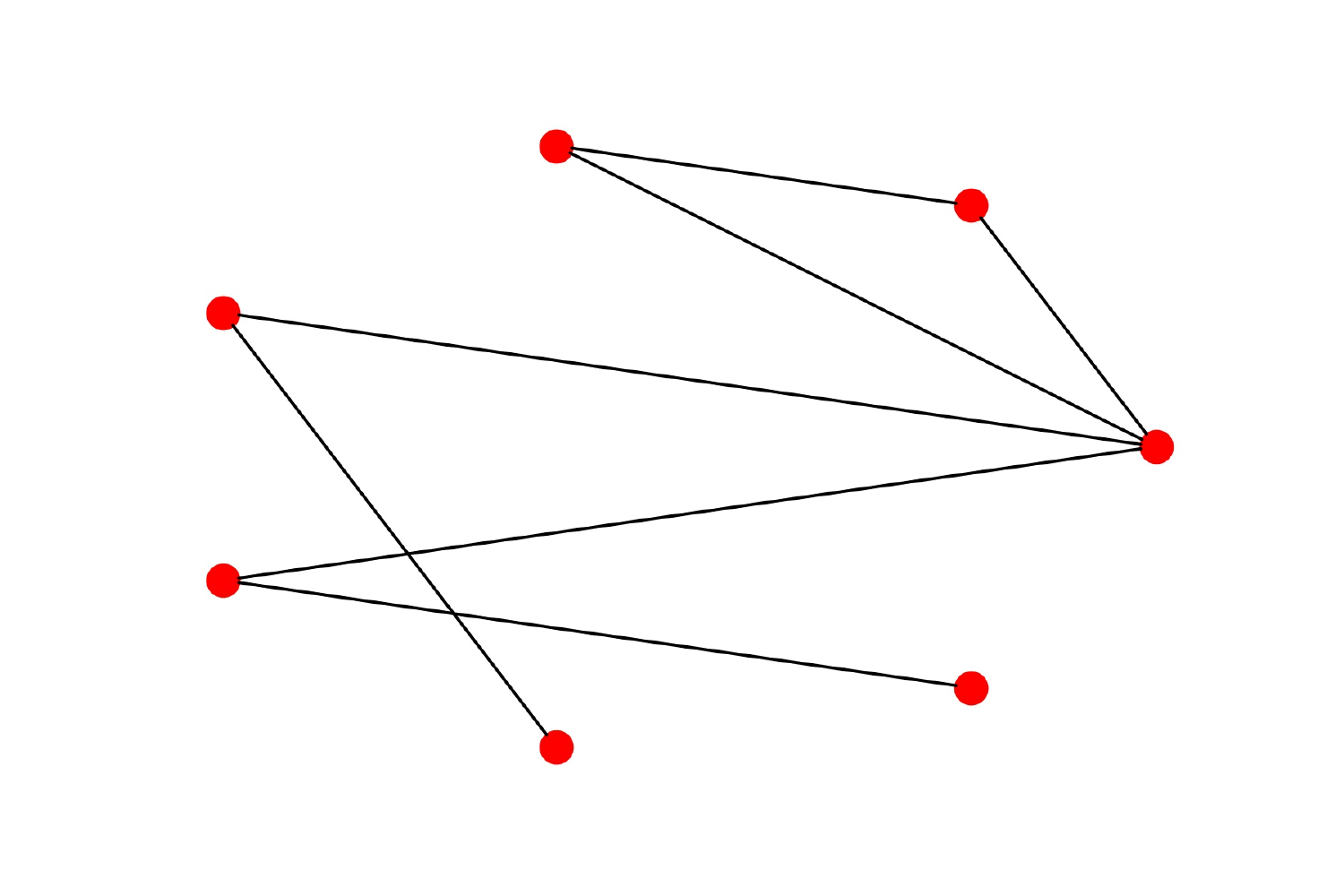} & 3 & 1 & 1 \\
        \hline
        \end{tabular}

    \caption{Two paths selected by the EFPs in the $7^\text{th}$ and $8^\text{th}$ iteration}
    \label{table:7_8_iteration}
\end{table}
      
\subsection{Physical interpretation of the selected features}

The selected Energy Flow Polynomials can be used to gain physical insight for the case of top tagging. Shown in Tables~\ref{tbl:efps} and \ref{table:7_8_iteration}  are the graphs, chromatic numbers $c$, $(\kappa,\beta)$ values, and cumulative $R_{30}$ values of the first eight EFPs selected by DisCo-FFS. We see that 5 of the first 6 EFPs selected are EFPs with $c=3$. A chromatic number of a graph is the number of colours one can put to the nodes, so that no edges are connected by the same colour. As noted in \cite{efp}, the chromatic number of an EFP is also a proxy for the number of prongs in the jet. In other words, $c=3$ EFPs are probes of 3-prong substructure -- exactly what one would expect to be relevant for top tagging.

Interestingly, there is one $c=2$ EFP selected in the first six EFPs. This probe of 2-prong substructure could be related to the two prongs consisting of the $b$-quark and the boosted $W$-jet inside the top quark. 

We also see from Table~\ref{tbl:efps} that both IRC-safe and unsafe probes of 3-prong substructure are useful for tagging. The first two EFPs have $\kappa=2$, and hence are an IRC-unsafe probe of hard radiation, with the first one being a 3-point correlator, and second one being a 4-point correlator.\footnote{We emphasize that all the HLFs we use in this work are actually IRC-safe in the end, since they are constructed from detector-reconstructed particles.}  IRC-safe EFPs ($\kappa=1$) are not selected until the fourth and fifth iteration.

In the seventh and eighth iterations, there appear to be two possible paths for the FS algorithm to take, i.e.\ two unique possibilities for the pairs of EFPs selected. These are shown in Table \ref{table:7_8_iteration}. 
In one of the paths, two IRC-unsafe EFPs probing the 2-prong substructure are selected with one of them probing small-angle radiation ($\beta=0.5$), and the other one probing hard/wide-angle radiation ($\beta=2$), which actually marks the first selected feature that probes wide-angle radiation. In the other path, we see the appearance of the first EFP which probes 4-prong substructure with small-angle radiation ($\beta=0.5$), and this is followed up by an IRC-safe EFP probing 3-prong substructure. 

Interestingly in our single run of \texttt{LorentzNet}-guided DisCo-FFS, the first 6 features are the same as Table \ref{tbl:efps}, whereas after that the $7^\text{th}$-EFP is the same one selected in Path 1 in \ref{table:7_8_iteration}. This confirms that the similar performance between DisCo-FFS with truth and with \texttt{LorentzNet} is no coincidence, and is likely because \texttt{LorentzNet} (being so high-performing) is quite close to the truth labels.

\section{Conclusions}

\label{section_conclusions}

In this work, we have introduced a new forward feature selection method, based on the distance correlation measure of statistical dependence --- dubbed DisCo-FFS. Our method can operate equally well on either truth-labels (for ab initio feature selection) or on the outputs of a pre-trained classifier (for explaining a ``black box" AI).

We demonstrated the performance of our method using the task of boosted top tagging, as boosted top jets have a rich substructure and many subtle correlations that have proven to be a fruitful laboratory for developing increasingly powerful state-of-the-art taggers in the HEP literature.

Following \cite{ado}, we have trained our DisCo-FFS method on a large set (7,000+) of Energy Flow Polynomials, which aim to provide a complete description of the jet substructure. We have seen that DisCo-FFS is very effective at selecting EFPs from this large feature set; DisCo-FFS can achieve nearly-state-of-the-art top tagging performance (matching that of ParticleNet-lite \cite{particlenet}) with a selection of just a small number of EFPs (less than 10). We also show how it outperforms the DO-ADO-FFS method of \cite{ado} (which we have attempted to replicate as closely as possible), consistently achieving higher tagging performance after each EFP that is selected.

The fact that our method falls short of the most state of the art deep learning methods (ParT~\cite{part}, PELICAN~\cite{pelican}, and LorentzNet \cite{lorenznet}) is interesting. Either our method is not fully optimal at selecting the features, or the 7,000+ EFPs we used as the basis of our study do not capture all the physics underlying top tagging. 
A possible follow-up study to further probe this question would be to
supplement the  7,000+ EFPs with additional jet substructure variables, for instance the subjettiness variables of \cite{dutta_landscape,Datta:2019ndh},
jet spectra and morphological features of~\cite{Chakraborty:2019imr,Chakraborty:2020yfc,Lim:2020igi},
or Boost Invariant Polynomials~\cite{bip}. 
This observation also raises the possibility that there might be more meaningful jet substructure variables out there, beyond those that are presently known, waiting to be discovered. This is obviously an interesting avenue for future research. 

Beyond simple object tagging, DisCo-FFS might also be able to shine for tasks --- such as building supervised classifiers for new physics discovery --- where calibration of the algorithm is difficult and a small number of well-understood features is preferable.
While particle physics is in an especially good position due to the presence of well-motivated bases of features (such as the used EFPs) such decompositions also exists for other domains, e.g. in the forms of wavelets applied to images (e.g. building on \cite{Rentala:2014bxa}).

In general, EFPs selected could make for a very lightweight and performant top tagger. This could have important applications to triggering~\cite{Duarte:2018ite}. For that,
a fast way to calculate EFPs on FPGAs would be required. Such will be interesting to explore further.

It would also be potentially illuminating to study the robustness of the selected EFPs under domain shift. For example, recently ATLAS released an official top tagging dataset~\cite{ATL-PHYS-PUB-2022-039}. One could compare the EFPs selected by DisCo-FFS on the different top tagging datasets, and see how one set of EFPs performs on the other dataset. 
One could also imagine training this method on a restricted set of HLFs (EFPs or otherwise) that are deemed to be ``well-modeled" by simulations. This could help with the calibration and robustness of taggers developed using simulation and deployed on data.

Overall, we observe the start of a positive feedback loop between deep learning method development and physics-motivated feature discovery.
Each one drives the other. Early top taggers~\cite{Adams:2015hiv} started with jet substructure variables like $N$-subjettiness. Then it looked like deep learning was able to go way beyond HLFs and we would have to rely on fully-automated feature engineering. Now there is some signs that we are coming full circle. Ultimately we may hope to match the performance of the SOTA deep learning taggers with just a handful of (yet-to-be-invented?) HLFs. This would be a very satisfying outcome, proving that deep learning doesn't have to be a black box but can drive fundamental  physics discoveries.

\section*{acknowledgements}

We are grateful to Taylor Faucett, Daniel Whiteson and especially Jesse Thaler for discussions and help regarding the $W$-jets dataset, the DO-ADO method, and truth vs.\ black-box guiding. We are also grateful to  Sitian Qian for assisting us with the \texttt{LorentzNet} classifier output. We thank Purvasha Chakravarti and Jose M. Muñoz Arias for helpful discussions. Finally, we thank Daniel Whiteson for comments on the draft.
GK acknowledges support by the Deutsche Forschungsgemeinschaft under Germany’s Excellence Strategy – EXC 2121  Quantum Universe – 390833306. The work of RD and DS was supported by DOE grant DOE-SC0010008. The authors acknowledge the Office of Advanced Research Computing (OARC) at Rutgers, The State University of New Jersey \url{https://it.rutgers.edu/oarc} for providing access to the Amarel cluster and associated research computing resources that have contributed to the results reported here.

\appendix

\begin{figure*}[]
    \centering
    \includegraphics[width=0.48\textwidth]{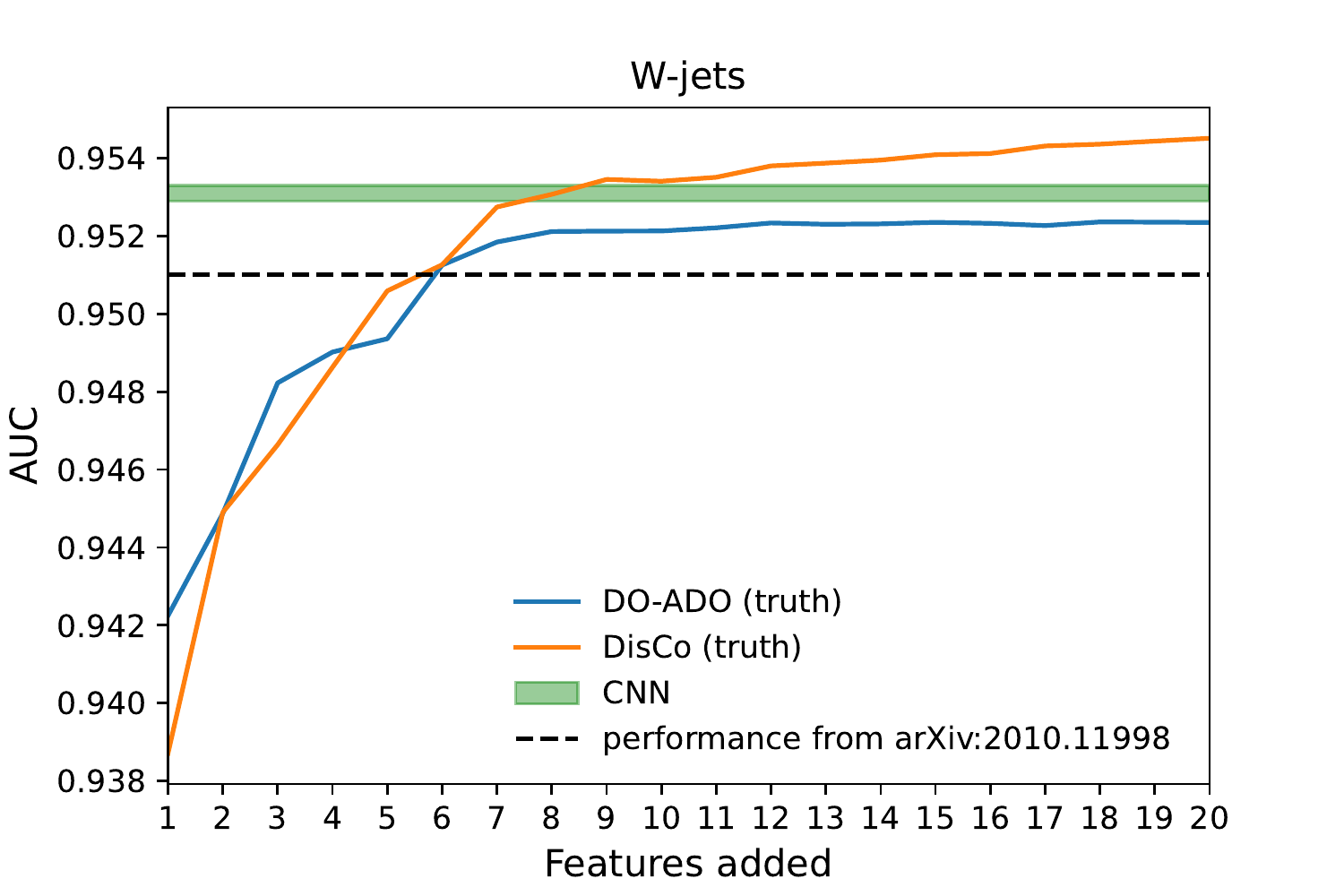}
    \includegraphics[width=0.48\textwidth]{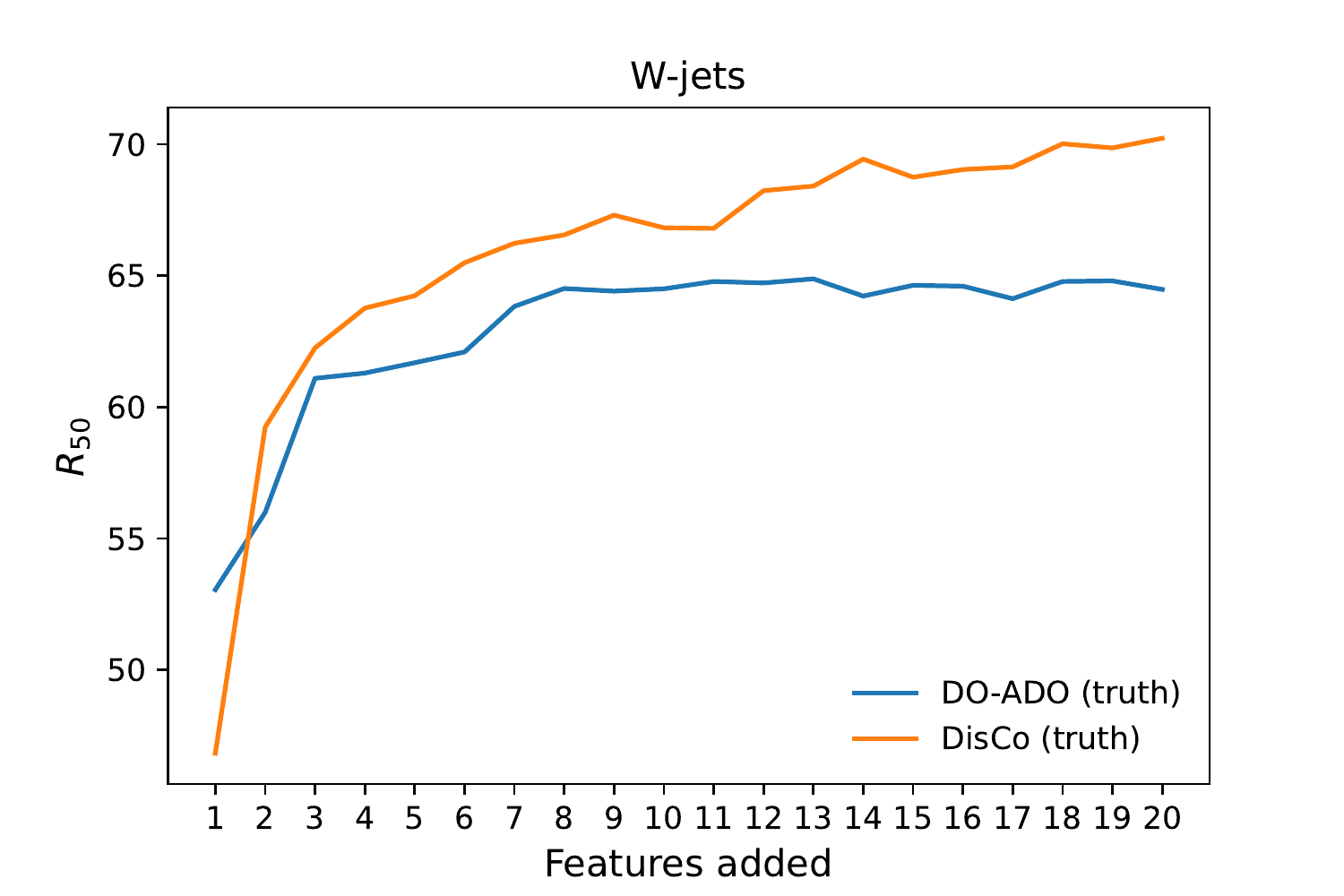}
\caption{Left: AUC vs.\ number of features selected, for DO-ADO (blue) and DisCo (orange), both truth-guided. The green line indicates the AUC of the reference CNN tagger from \cite{ado}, while the black dashed line indicates the performance that truth-guided DO-ADO achieved in \cite{ado}. Here we see our version of the truth-guided DO-ADO method saturates at a slightly higher AUC of 0.952 (but still short of the CNN AUC), whereas the DisCo-FFS method reaches the CNN AUC after 8 features, and is able to exceed the CNN AUC. Right: Same comparison but in terms of the $R_{50}$ (rejection power at 50\% tpr) metric. }
\label{fig:wjets}
\end{figure*}

\section{Validation of our implementation of DO-ADO-FFS}\label{appendix:ado}

\subsection{The DO-ADO feature selection method}

In this appendix, we validate our implementation of the DO-ADO feature selection method of \cite{ado}. This method is based on the {\it decision ordering} (DO) and {\it average decision ordering} (ADO) metrics, which we will now explain. 

For a signal event $x_s$ and a background event $x_b$, the DO metric is given by
    \bea\label{eq:do}
    & {\rm DO}(x_s,x_b;y_{\rm pred},y_{\rm ref})   =  \\  & \Theta\Big((y_{\rm pred} (x_s)- y_{\rm pred}(x_b))\times
         (y_{\rm ref}(x_s)- y_{\rm ref}(x_b))\Big),  
    \eea

\noindent
where $\Theta$ is the Heaviside step function. In other words, ${\rm DO}=1$ (${\rm DO}=0$) if the pair of events has the same (different) ordering under $y_\text{pred}$ as under the reference classifier $y_\text{ref}$.

Meanwhile, Average Decision Ordering is defined over a dataset ${\mathcal D}$ consisting of pairs of signal and background events:
\beq\label{eq:ado}
{\rm ADO}({\mathcal D};y_{\rm pred},y_{\rm ref})=\left\langle {\rm DO}(x_s,x_b;y_{\rm pred},y_{\rm ref})\right\rangle_{(x_s,x_b)\sim {\mathcal D}}
\eeq
In other words, ADO is the average of the DO metric over the dataset. 

The DO-ADO feature selection algorithm \cite{ado} also follows the same steps 1 and 4, as described in section \ref{section:general_method}. For steps 2 and 3, we have
\begin{enumerate}[label=Step \arabic*:,start=2, align=left, leftmargin=*]
    \item The confusion set $X_0$ is formed out of pairs of (signal,background) events with $\text{DO} (x_s,x_b;y_\text{pred},y_\text{ref})=0$. It is too computationally intensive to find and analyze all possible pairs of events with $\text{DO}=0$, so only a randomly selected subset of (signal,background) pairs is considered for $X_0$. 

    \item The relevance score for each feature $f$ is defined as 
    \beq 
    s_f = \text{ADO}(X_0;f,y_\text{ref}).
    \eeq
    So a feature with a larger ADO value would be one for which more events in the confusion set are correctly ordered by the feature. The idea of DO-ADO-FFS is to identify the feature at every step that most correctly orders signal vs.\ background events that are incorrectly ordered by the previous step, with respect to the the reference classifier $y_\text{ref}$. 
\end{enumerate}
    
\subsection{Validation with $W$-tagging}

To validate our implementation of DO-ADO-FFS, we train it on the same $W$-tagging dataset considered in \cite{ado} with respect to truth labels,\footnote{We could not perform DO-ADO-FFS with respect to the pre-trained CNN because this was not made publicly available at the time of this publication.}  and demonstrate that we achieve the same performance as shown there. 

As in \cite{ado}, we start with an initial feature set of
\beq
{\mathcal F}_{initial}={\mathcal F}_2= \{m_J,p_T\}
\eeq
Here we apply both truth-guided DO-ADO-FFS and DisCo-FFS to the same set of EFPs considered in \cite{ado} and this paper. The results (as AUC and $R_{50}$ vs number of features selected) are shown in Fig.~\ref{fig:wjets}, together with the performance metrics for a reference CNN tagger from \cite{ado}, as well as the reference AUC value of 0.951, at which the truth-guided ADO in \cite{ado} was mentioned to saturate after 7 features.

For the ADO method, we see that the AUC reaches around 0.951 after 7 features. This matches the description in \cite{ado} and demonstrates that we have successfully validated the implementation of DO-ADO-FFS. Interestingly, however, we notice that the AUC of our version saturates at a slightly higher AUC of around 0.952. 

Meanwhile, Disco-FFS again outperforms DO-ADO-FFS: it reaches the CNN AUC after 8 features, and actually proceeds to {\it exceed} the performance of the CNN -- all without using any knowledge of the CNN classifier output! This shows the potential promise of a well-designed forward feature selection method operating on a well-chosen feature set: it could conceivably show that a deep learning classifier is not actually as state-of-the-art as previously thought.

\section{Hyperparameters and Architectures}\label{section:hyperparams}

For our feature set, we use $\log (\text{EFP}+10^{-40})$,  instead of the bare EFPs as our features, during training, as well as during feature selection, and we see that this leads to a better performance. 

For DisCo-FFS, we use events in $0.3<y_\text{pred}<0.7$ as our confusion set $X_0$. The boundaries of this window are important hyperparameters of our algorithm, and we settled on this choice after scanning through different window sizes and seeing where the performance of the method was best.

Due to computational constraints, we actually calculate DisCo using minibatches. We divide the confusion set $X_0$ into minibatch sizes of 2048, and then average over all the minibatches to estimate DisCo over the confusion set.

\texttt{Tensorflow} was used for training classifiers for DisCo-FFS and DO-ADO-FFS, and the following hyperparameters were used: 
\begin{itemize}

    \item 2 hidden layers of 16 nodes with \text{ReLU} activation, final output layer with \texttt{softmax} activation.
    \item A  \texttt{RobustScaler} is fitted on the training and validation data combined and is used to rescale the dataset
    \item We use the \texttt{Adam} optimizer with default hyper-parameters for 500 epochs, with mini-batch size $=512$. 
    \item Model checkpoint is used to save the model with the minimum validation loss.
    
\end{itemize}

We observed that the final $R_{30}$'s were higher after the use of a slightly bigger network with $ 32 \times 32 $ hidden layers, so we retrained all the features (after the FFS) with this network, and obtained our final $R_{30}$'s, including Fig.~\ref{fig:performance}, with this network. 

The DNN trained on all 7k EFPs uses the same hyper-parameters as discussed above, but we use a slightly bigger network with 3 hidden layers of 32 nodes.

For both the truth-guided DisCo-FFS and DO-ADO methods, we apply feature selection to the combined training and validation sets. However for the \texttt{LorentzNet}-guided versions, we apply the feature selection only to the validation set. This is because we noticed a significant overfitting of \texttt{LorentzNet} to the training set, as compared to the validation and the test set. 

\section{Affine-Invariant Distance Correlation}\label{section:disco}

Distance Correlation (DisCo), is a correlation metric which can quantify non-linear correlations in the joint distribution of two random vectors $(\vec X,\vec Y)$ of arbitrary dimension \cite{disco,disco_1,disco_2,disco_3}. In particular, DisCo is zero iff $\vec X$ and $\vec Y$ are statistically independent ($p(\vec X,\vec Y)=p(\vec X)p(\vec Y)$), and positive otherwise.

With $\vec X$ and $\vec Y$ as 1-D vectors, DisCo has used been previously used in physics for decorrelation of neural networks against mass \cite{DiscoFever}. However, DisCo is even more powerful than that -- it can also measure statistical dependence of {\it multivariate} distributions, a powerful property that enables the forward feature selection algorithm described in this work.

For our case, $\vec X = y_\text{truth}$ is a 1-D vector, and $\vec Y = (f_{i_1} , f_{i_2},\dots,f_{i_n})$ is an $n$-dimensional feature vector. The population value of squared distance covariance of $\vec X$ and $\vec Y$ is given by
\begin{equation}
    {\begin{aligned}\operatorname {dCov} ^{2}(\vec X,\vec Y):={}&\operatorname {E} [\|\vec X-\vec X'\|\,\|\vec Y-\vec Y'\|]+\\& \operatorname {E} [\|\vec X-\vec X'\|]\,\operatorname {E} [\|\vec Y-\vec Y'\|]\\ &-2\operatorname {E} [\|\vec X-\vec X'\|\,\|\vec Y-\vec Y''\|].\end{aligned}}
\end{equation}
Distance correlation is given by
\begin{equation}
    \operatorname {dCor} ^{2}(\vec X,\vec Y)={\frac {\operatorname {dCov} ^{2}(\vec X,\vec Y)}{\sqrt {\operatorname {dCov} ^{2}(\vec X,\vec X)\,\operatorname {dCov} ^{2}(\vec Y,\vec Y)}}},
\end{equation}
 which is normalized between 0 and 1. 
 
 Finally, using the covariance matrices $\Sigma_X$, $\Sigma_Y$, affine-invariant distance correlation is simply
 \begin{equation}
    {\operatorname{\overline{{dCor}}}^2} (\vec X,\vec Y) = \operatorname{dCor}^2(\Sigma_X^{-1/2}\vec  X,\Sigma_Y^{-1/2}\vec Y).
 \end{equation}
In this work, we use the \texttt{dcor} package \cite{dcor} for the computation of distance correlation and affine-invariant distance correlation.

\bibliography{main}

\end{document}